\documentclass[11pt]{article}
\usepackage{aas_macros,amsmath,amssymb,comment,cite,esint,graphicx,mathtools}
\usepackage[margin=.8in,letterpaper]{geometry}
\usepackage[colorlinks=true]{hyperref}
\usepackage[affil-it]{authblk}
\usepackage{subcaption}
\usepackage[utf8]{inputenc}
\usepackage{mathrsfs}
\usepackage{appendix}
\usepackage{amssymb}
\usepackage{float}
\usepackage{color}
\usepackage{cite}
\usepackage{hyperref}
\usepackage{longtable}
\hypersetup{pageanchor=false}
\usepackage{indentfirst}
\usepackage{url}
\usepackage{float}
\usepackage{caption}
\usepackage[numbers,square,comma,sort&compress,merge]{natbib}
\usepackage{esint}
\usepackage{overpic}
\usepackage{graphicx}
\usepackage{epsf,amsmath,bbold,amsfonts,stmaryrd}
\usepackage{textcomp}
\usepackage{ulem}
\usepackage{tikz}
\usepackage{multirow}

\numberwithin{equation}{section}
\let\originalleft\left
\let\originalright\right
\renewcommand{\left}{\mathopen{}\mathclose\bgroup\originalleft}
\renewcommand{\right}{\aftergroup\egroup\originalright}
\mathcode`\*="8000
{\catcode`\*=\active\gdef*{\mathclose{}\,\mathopen{}}}

\def\d{\delta}
\def\D{\Delta}
\def\f{\frac}

\def\lm{\lambda}

\def\m{\mu}
\def\n{\nu}
\def\nn{\nonumber}

\def\t{\theta}
\def\T{\Theta}

\newcommand{\mE}{{\mathcal E}}
\newcommand{\be}{\begin{equation}}
\newcommand{\ee}{\end{equation}}
\newcommand{\bea}{\setlength\arraycolsep{2pt} \begin{eqnarray}}
\newcommand{\eea}{\end{eqnarray}}
\newcommand{\mm}{\mathrm}
\newcommand{\mc}{\mathcal}

\def\fft#1#2{{\frac{#1}{#2}}}

\begin{document}

\title{Imaging a Semi-Analytical Jet model Generated by 3D GRMHD Simulation}

\author{
Ye Shen$^{1}$, Yehui Hou$^{1}$, Zhong-Ying Fan$^{2}$, Minyong Guo$^{3}$,
Bin Chen$^{1,4,5\ast}$}
\date{}

\maketitle

\vspace{-10mm}

\begin{center}
{\it
$^1$Department of Physics, Peking University, No.5 Yiheyuan Rd, Beijing
100871, P.R. China\\\vspace{4mm}

$^2$ School of Physics and Material Science,
 Guangzhou University, Guangzhou 510006, P.R. China \\\vspace{4mm}

$^3$ Department of Physics, Beijing Normal University,
Beijing 100875, P. R. China\\\vspace{4mm}

$^4$Center for High Energy Physics, Peking University,
No.5 Yiheyuan Rd, Beijing 100871, P. R. China\\\vspace{4mm}

$^5$ Collaborative Innovation Center of Quantum Matter,
No.5 Yiheyuan Rd, Beijing 100871, P. R. China\\\vspace{2mm}
}
\end{center}

\vspace{8mm}

\begin{abstract}

Employing 3D GRMHD simulation, we study the images of a geometrically thin jet, whose emissions concentrate on its surface, for accretion system surrounding a central spinning BH. By introducing a strong magnetic field, we observe three phases of BH accretion evolution: (a) initially, both the accretion rate and the magnetic flux on the horizon gradually increase; (b) at an intermediate stage, the magnetic flux approximately reaches saturation, and a jet forms via the Blandford-Znajek (BZ) mechanism; (c) ultimately, the entire system achieves a dynamic equilibrium, and a magnetically arrested disk (MAD) forms. We carefully study the jet images during the saturation and MAD regimes at various frequencies and from different observational angles. We reveal the presence of U-shaped brighter lines near the jet surface boundaries, which can be attributed to the photons whose trajectories skim over the jet surface. The existence of these brighter lines is a unique feature of a geometrically thin jet. Moreover, we notice that the jet images are  relatively insensitive  to the observed frequencies of interest. Additionally, we observe that the time-averaged images for the highly oscillating MAD regime show only slight differences from those of the saturation regime.


\end{abstract}

\vfill{\footnotesize $\ast$ Corresponding author: bchen01@pku.edu.cn}

\maketitle

\newpage
\baselineskip 18pt

\section{Introduction}

To date, the Event Horizon Telescope (EHT) Collaboration has released a series of images of black holes (BHs) taken by very long baseline interferometry (VLBI) observations. These images include the ones  of the BHs located at the centers of M87* \cite{EventHorizonTelescope:2019dse} and Sgr A \cite{EventHorizonTelescope:2022wkp} captured at 230 GHz, a polarized image of M87* at 230 GHz \cite{EventHorizonTelescope:2021srq}, and an image of M87* at 86 GHz \cite{Lu:2023bbn}. All of these images consistently exhibit a ring-like structure, which is interpreted as the result of gravitational lensing of millimeter-wave emissions surrounding the event horizons \cite{EventHorizonTelescope:2019pgp}. In accordance with active galactic nuclei (AGN), millimeter-wave emissions are believed to originate from accretion flows and jets. Undoubtedly, the final representation in black hole photography relies not only on the strong gravitational lensing of the black hole but also significantly on the intrinsic properties of the light source itself. Therefore, to comprehend black hole images, it is crucial to understand not only the nature of black holes themselves but also the mechanisms of accretion flows and jets.

Numerous studies have been conducted to understand the impact of different accretion flows on black hole imaging. These primarily include spherical accretions \cite{Narayan:2019imo}, geometrically thin disks \cite{Gralla:2019xty}, geometrically thick disks \cite{Vincent:2022fwj}, and numerical General Relativistic Magnetohydrodynamic (GRMHD) models \cite{Chatterjee2020}. However, research on black hole imaging using jets as the emission source is comparatively less explored \cite{Davelaar:2023dhl}, largely due to the indeterminacy of the initial conditions required to generate a jet in numerical simulations. Besides, there is a notable lack of research concerning the imaging of jets at the scale of a black hole's event horizon.

The investigation of jet imaging is important for understanding accretion systems that incorporate both accretion disks and jets. Previous studies have attempted to image the entire accretion system (including the disk, corona, and funnel region) generated from GRMHD simulations \cite{PATOKA, Chatterjee2020}. In these studies, the emission from the jet could not be distinguished from that of the entire system. Elementary works on jet images have been conducted based on large-scale ($10^3-10^4\,r_g$) MHD models \cite{Miodusz1997, Curd2022}, but the gravitational effect is hardly discernible in the images as it is significantly weaker than other effects. Recently, the polarized images of jets on a scale of $10-10^2\,r_g$ have been explored by combining GRMHD and GR ray-tracing (GRRT) \cite{Tsunetoe2020, synchrotron2023}. These studies are based on complex synchrotron emission models and are numerically demanding. However  the relationship between jet structures and the images remains almost unexplored, as far as we know. In addition, analytical models of jets have also been studied, such as the semi-empirical magnetically dominated model \cite{Broderick_model_2009, Broderick_model_2020} and the simplified dual-cone model \cite{Papoutsis:2022kzp}. These models have facilitated concrete analyses of jet structures and characteristics of jet images. However, as with other analytical models of accretion flow, ideal conditions must be imposed to render the systems solvable, and the chaotic time evolution has barely been considered.

In this work, by employing 3D GRMHD simulation, we study the horizon-scale ($\lesssim 10\, r_g$) images of jet generated by initially imposing a strong magnetic field. We explore the features of the jet in different epochs, as well as the relations between the images and jet structures. It has been proposed theoretically that jets can exhibit either thick or thin characteristics in terms of their geometric and optical properties \cite{optthin_jet,optthick_jet}. The dominant radiation mechanisms within various accretion systems can generate distinct images. Here, we focus specifically on the geometrically thin relativistic jets. We develop a simplified emission model, particularly suitable for geometrically thin emission sources, to significantly mitigate the computational cost associated with solving the radiative transfer equation. We manage to identify unique characteristics belonging to the images of geometrically thin jets.

The paper is organized as follows. In Section \ref{sec:emi_model}, we introduce the simple emission model for geometrically thin jets, neglecting self-absorption effects for thermal synchrotron radiation. This approximation is valid within the observed frequency range of interest (40-350 GHz). In Section \ref{sec3}, we provide a brief introduction to the basic setup for 3D GRMHD simulations. In Section \ref{sec4}, we present the main numerical results, including three epochs representing the evolution of the entire accretion system, as well as the images of the jet at various frequencies and observation angles. Finally, in Section \ref{sec5}, we discuss future directions for research.

\section{Emission Model}
\label{sec:emi_model}

\begin{figure}
    \centering
    \includegraphics[width=0.45\linewidth]{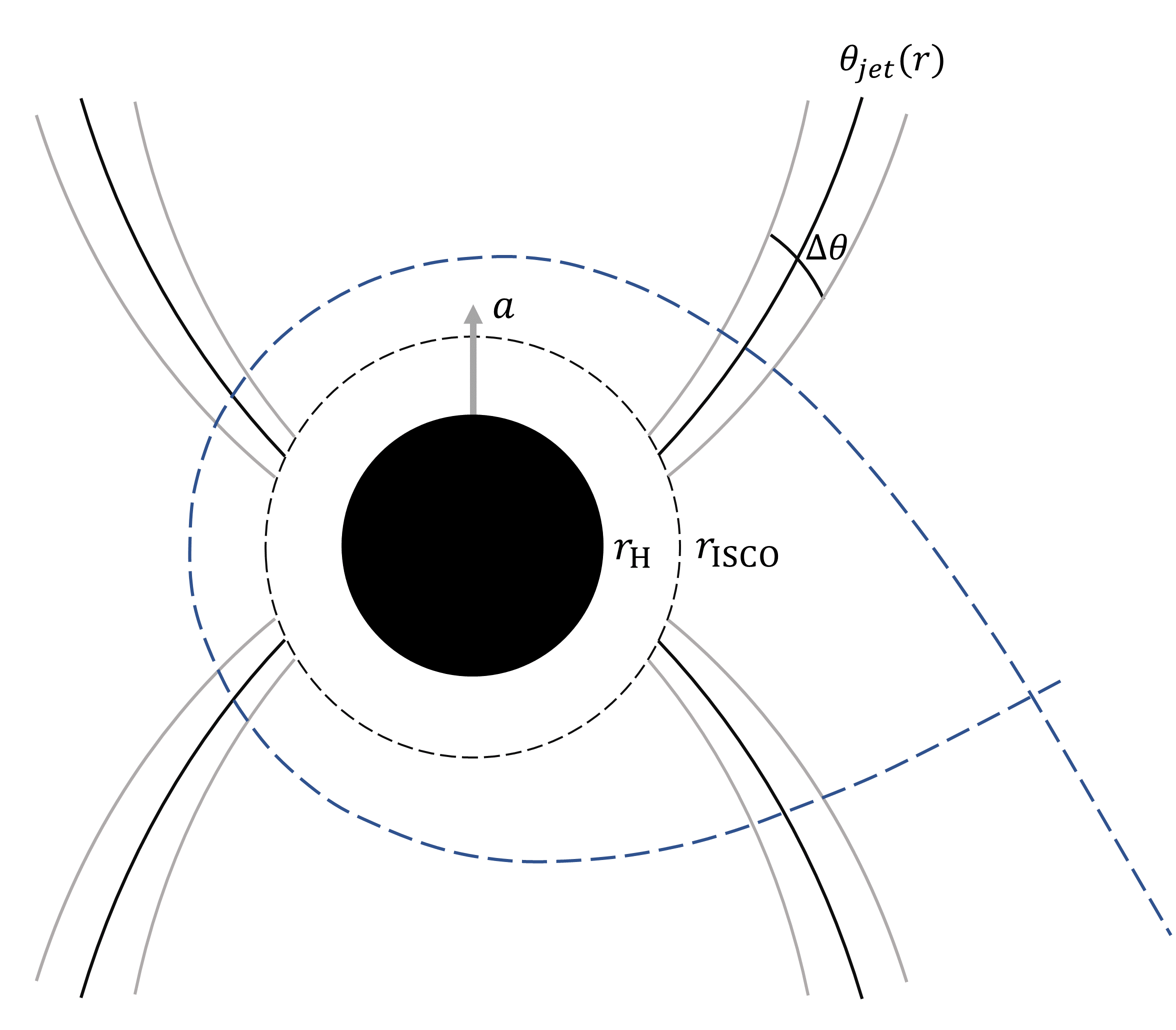}
    \caption{Scheme of the geometrically thin jet model. The jet surface is defined by $\t = \t_{jet}(r)$, and the thickness is given by $ \t_{jet}(r) \pm \D\t(r)/2$. }
    \label{fig:jet_rad}
\end{figure}

We focus on a geometrically thin jet model in which the ratio of magnetic pressure to gas pressure reaches its maximum on the surface. When the gas pressure is much lower than the magnetic pressure, or equivalently, when the sound speed ($c^2_s\sim \frac{p}{\rho}$) is much lower than the Alfv{\'e}n velocity ($c^2_A\sim \frac{b^2}{\rho}$), the interactions among fluid particles can be ignored when considering the emission. The fluid can be treated as an ideal gas affected by a strong magnetic field. In this case, the emission is primarily distributed on the surface of the jet rather than in the interior of the funnel region. See Fig.~\ref{fig:jet_rad} for an illustration. We will revisit this point when discussing the numerical results.

Under the influence of a strong electromagnetic field, synchrotron radiation predominates in the emission of photons by electrons. The power spectrum for a single electron, as observed in the fluid rest frame, can be expressed as follows \cite{1979Lightman} 
\be P_\nu=\sqrt{3}\fft{e^3}{m_e}\mc{E} F\Big( \fft{\nu}{\nu_c}\Big)\,,\quad F(x)=x\int_x^\infty K_{5/3}(x')dx' \,,\ee
where $\nu$ is the emission frequency, $\nu_c=3e\gamma^2 \mc{E}/4\pi m_e$ is the charateristic frequency ($\gamma$ is the Lorentz factor of electrons) and
\bea
    \mE = \sqrt{b^2 - \f{(k^{\m}b_{\m})^2}{\n^2}} \,,
\eea
recodes the radius curvature of cyclotron electrons, where $k^\mu$ is the wave $4$-vector. For ultra-relativistic electrons, the radiation is highly concentrated in the directions of motion. Hence the single electron emissivity can be well approximated by $J_e\approx P_\nu \delta(\Omega-\Omega_{\vec{v}})$, where $\Omega$ ($\Omega_{\vec v}$) stands for the radiation (electron-moving) direction. Taking the distribution of electrons into account, the total emissivity is given by
\be j_\nu=\fft{1}{4\pi}\int d\gamma d\Omega_{\vec v}\fft{dn}{d\gamma}\,J_e \,,\ee
depending on which kind of electrons, thermal or non-thermal, dominates in the radiation. If thermal electrons dominates, the distribution function $dn/d\gamma$ is the Maxwell-J{\"{u}}ttner distribution. For higher frequency $\nu\gg e\T_e^2\mE/9\pi m_e$ or $X\gg 1$, the emissivity takes the form of \cite{leung2011numerical}
\bea
	j_{\n} \approx \f{\sqrt{2}\pi  e^2}{6\T_e^2} n \nu f(X) e^{-X^{1/3}}\,, \quad X =  \f{9\pi m_e \n}{e\T_e^2\mE}  \,,
\eea
where $\T_e = k_BT_e/m_e$ is a dimensionless temperature of electrons and
\be f(X)=\big[  1 + 2^{11/12} X^{-1/3} \big]^2 \,, \ee
is a fitting function in our numerical code. If non-thermal electrons dominate in synchrotron radiation, the distribution function is expected to adhere to a single or broken power-law distribution. However, when we concentrate on the images of our model, the outcomes produced by thermally and non-thermally distributed electrons appear to be virtually indistinguishable, consistent with the results in \cite{synchrotron2023}. Therefore, in this study, we opt to disregard the non-thermal case.


Given the emission profile, we will employ the GRRT technique to identify the light rays emitted from the jet surface. Subsequently, we will integrate the radiative transfer equation along each light ray to derive the observed intensity. The radiative transfer equation, in the absence of scattering, is as follows
\be \fft{d}{d\lambda}\Big( \fft{I_\nu}{\nu^3}\Big)=\fft{j_\nu-\alpha_\nu I_\nu}{\nu^2} \,,\label{transport}\ee
where $\lambda$ represents the affine parameter and $\alpha_\nu$ is the absorption coefficient. At high frequencies, $\alpha_\nu$ becomes exponentially small, and can thus be safely ignored. This is a valid approximation within the range of observed frequencies we are interested in, specifically between 40 and 350 GHz. Once the light rays are determined (denoted by $x^\mu(\lambda)$), the solution to Eq.(\ref{transport}) can be expressed as
\be I_{\nu_o}=\nu_o \int_{\lambda_i}^{\lambda_f} d\lambda\, g^2\big(x^\mu(\lambda),k^\mu(\lambda)\big)\,j_\nu\big(x^\mu(\lambda),k^\mu(\lambda)\big)  \,,\ee
where $\nu_o$ represents the frequency of the light reaching the observer, and $g(x^\mu,k^\mu)=\nu_o/\nu$ is the redshift factor, which encompasses both the Doppler shift and gravitational redshift. However, as we are only considering the jet surface as the emission source, the radiative transfer equation can be simplified and discretized as follows:
\begin{equation}
    I_{\n_o} = \n_o \sum_{i=1}^N \d\lm_i \, g_i^2 j_{\n,i} \,
    \label{eq:emission}
\end{equation}
where $i$ denotes the number of instances that the null geodesic crosses the jet surface, while $N\in {\bf N^+}$ represents the maximum crossing times, as determined by pure gravitational lensing \cite{Gralla:2019xty}. The quantities $g_i, j_{\n,i}$ are evaluated exactly at the jet surface, whereas the factor $\d \lm_i$ is the change in the affine parameter when the light crosses the emission region. As depicted in Fig.~\ref{fig:jet_rad}, the thickness of the emission region in our model is small but finite (see section \ref{sec3} for details), which necessitates the numercial calculation of $\d\lm_i$.

\section{Source Generation}
\label{sec3}

We utilize \textbf{HARMPI}, an open-source GRMHD tool, to simulate the accretion process near a supermassive BH. It numerically solves the conservation equations and the Maxwell equations within an ideal magnetofluid context:
\begin{equation}
    \label{eq:conservation}
    \begin{array}{l}
    \nabla_{\mu}(\rho u^{\mu})=0\,,
    \\
    \nabla_{\mu}T^{\mu \nu}=0\,,
    \\
    \nabla_{\mu}{}^\ast{F}^{\mu \nu}=0\,,
    \end{array}
\end{equation}
through Godunov scheme \cite{HARM2003, HARM2006, godunov}. In the equations above, partial differential equations involving time are used to evolve the primary values of various quantities, the mass density $\rho$, the internal energy $\varepsilon$, the 3-velocity $u^{i}$, and the magnetic field $B^{i}$. These quantities completely describe the state of the accretion system, provided that the equation of state (EoS) for the plasma is given. The time component of the dual Maxwell equations ($\nabla\cdot\vec{B}=0$), being time-independent, serves as a no-monopole constraint for magnetic field. In ideal MHD, which adheres to the condition: $F_{\mu \nu}u^{\nu}=0$, it is useful to introduce the magnetic field measured in the fluid frame, defined as
\begin{equation}
    b^{\mu}=-{}^\ast F^{ \mu \nu}u_{\nu}=-\frac{1}{2}\epsilon^{\mu \nu \kappa \lambda}u_{\nu}F_{\kappa \lambda}\,.
\end{equation}
It is related to the 3-vector $B^{i}={}^{\ast}F^{ it}$ as
\begin{equation}
    b^t=g_{i\mu}B^{i}u^{\mu},
    ~~~~~~
    b^i=\frac{B^{i}+b^{t}u^{i}}{u^t}\,.
\end{equation}
Using $b^\mu$, the dual Maxwell tensor can be expressed as
\be     {}^{\ast}F^{\mu \nu}=b^{\mu}u^{\nu}-b^{\nu}u^{\mu}\,.\ee
It follows that the total stress-energy tensor is given by  
\bea\label{eq:tensor}
    &&T^{\mu\nu}=T^{\mu\nu}_{\mm{IF}}+T^{\mu\nu}_{\mm{EM}}\,,\\ 
    &&T^{\mu \nu}_{\mm{IF}}=(\rho+\varepsilon+p)u^{\mu}u^{\nu}+p g^{\mu \nu}\,,\\
    &&T^{\mu \nu}_{\mm{EM}}=(u^{\mu}u^{\nu}+\frac{1}{2}g^{\mu\nu})b^2-b^{\mu}b^{\nu}\,.
\eea
We opt for the monotonized central limiter to extropolate primary values on zone faces and the Lax-Friedrichs approximation to determine the fluxes crossing these zone faces (see \cite{toroBook}). The constrained transformation, introduced by T{\'o}th \cite{CT-Flux}, is employed to preserve the no-monopole condition. To ensure the regularity of the accretion flow on the event horizon, we work within the Kerr-Schild (KS) coordinates, in which the Kerr metric reads
\bea
        ds^2=&-&\left(1-\frac{2Mr}{\Sigma} \right)dt^2+\frac{4Mr}{\Sigma}dtdr+\left(1+\frac{2Mr}{\Sigma} \right)dr^2+\Sigma d\theta^2 \nn\\
        &+&\left(r^2+a^2+\frac{2Mr\,a^2\sin^2\theta}{\Sigma} \right) \sin^2\theta d\varphi^2  \nn\\
        &-&\frac{4Mar\sin^2\theta}{\Sigma}dt d\varphi 
        -2a \left(1+\frac{2Mr}{\Sigma} \right) \sin^2\theta dr d\varphi \,,
\eea
where $\Sigma=r^2+a^2\cos^2\theta$, $a$ is the BH spin per unit mass and $\Delta=r^{2}-2Mr+a^{2}$. 
For higher efficiency, we carry out numerical integrations in modified KS (MKS) coordinates $x^\mu$ \cite{HARM2004}, defined as
\begin{equation}
    \label{eq:KStoMKS}
    \begin{cases}
    t=x_{0},~~\varphi=x_{3}
    \\
    r=r_{0}e^{x_{1}}
    \\
    \theta=\pi x_{2}+\frac{1}{2}(1-h)\sin(2\pi x_{2})
    \end{cases}
\end{equation}
The grids are evenly incised in MKS coordinates. In the context of KS coordinates, this implies that more cells are generated near the event horizon, which is the focal region for gravitational lensing. The $h$ in Eq.~\eqref{eq:KStoMKS} is an index ranging from 0 to 1. A smaller $h$ yields more cells near the equatorial plane. To ensure sufficient cell coverage near the jet surface and inside the event horizon $r_H$, we set $h=0.7$ and $r_0=0.87r_{\rm H}$.

The resolution is: $N_{r}\times N_{\theta}\times N_{\phi}=128\times 64\times 32$. The 2D scheme for primary recovery, introduced in \cite{HARM2006}, is employed to update the primary values following numerical time integration at each step. In this study, we focus on a moderately rapidly rotating BH with $a/M=0.9$, and we set $M=1$. Consequently, we obtain the radius of the event horizon $r_{\rm H}\approx 1.44r_g$ and the innermost stable circular orbit (ISCO) $r_{\rm ISCO}\approx 2.32r_g$, where $r_g$ represents the gravitational radius. Initially, we introduce a geometrically thick accretion disk, which can be characterized by the torus model \cite{TorusFM}, with the inner boundary set to 15$r_{g}$ and center radius set to 35$r_{g}$. The atmospheric conditions outside the torus are defined as $\rho_{\rm atm}=10^{-5}(r/r_g)^{-3/2}$ and $\varepsilon_{\rm atm}=10^{-7}(r/r_g)^{-5/2}$, mirroring the distribution of the Bondi accretion \cite{frankBook}. We introduce a single magnetic field loop by defining the vector potential as per the method outlined in \cite{BHAC}:
\begin{equation}
    \label{v_potential}
    A_{\phi}\sim \mm{max}(\rho/\rho_{max}-0.2,0).
\end{equation}
We establish a strong magnetic field within the accretion system by setting $\beta=p_{max}/b^{2}_{max}=2$ initially. A substantial magnetic field can disrupt the stability of the disk \cite{MRI1991, MRI1998} and may utimately generate a relativistic jet through the BZ process \cite{SashaNote, BZ, jetformation, frankBook}.

\begin{figure}
    \centering
    \includegraphics[width=0.45\textwidth]{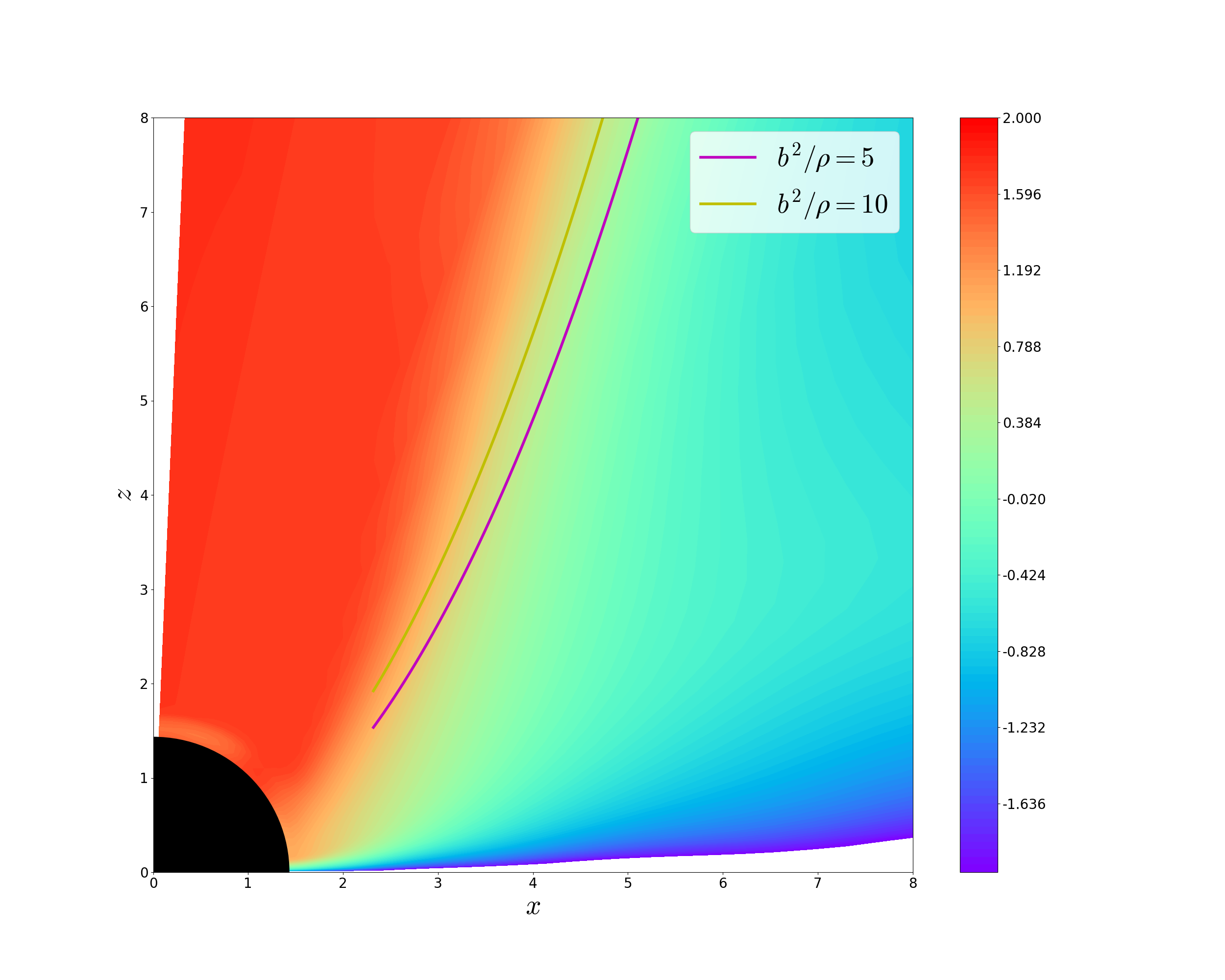}
    \includegraphics[width=0.45\textwidth]{beta2_distri_ave2-3.png}
    \includegraphics[width=0.45\textwidth]{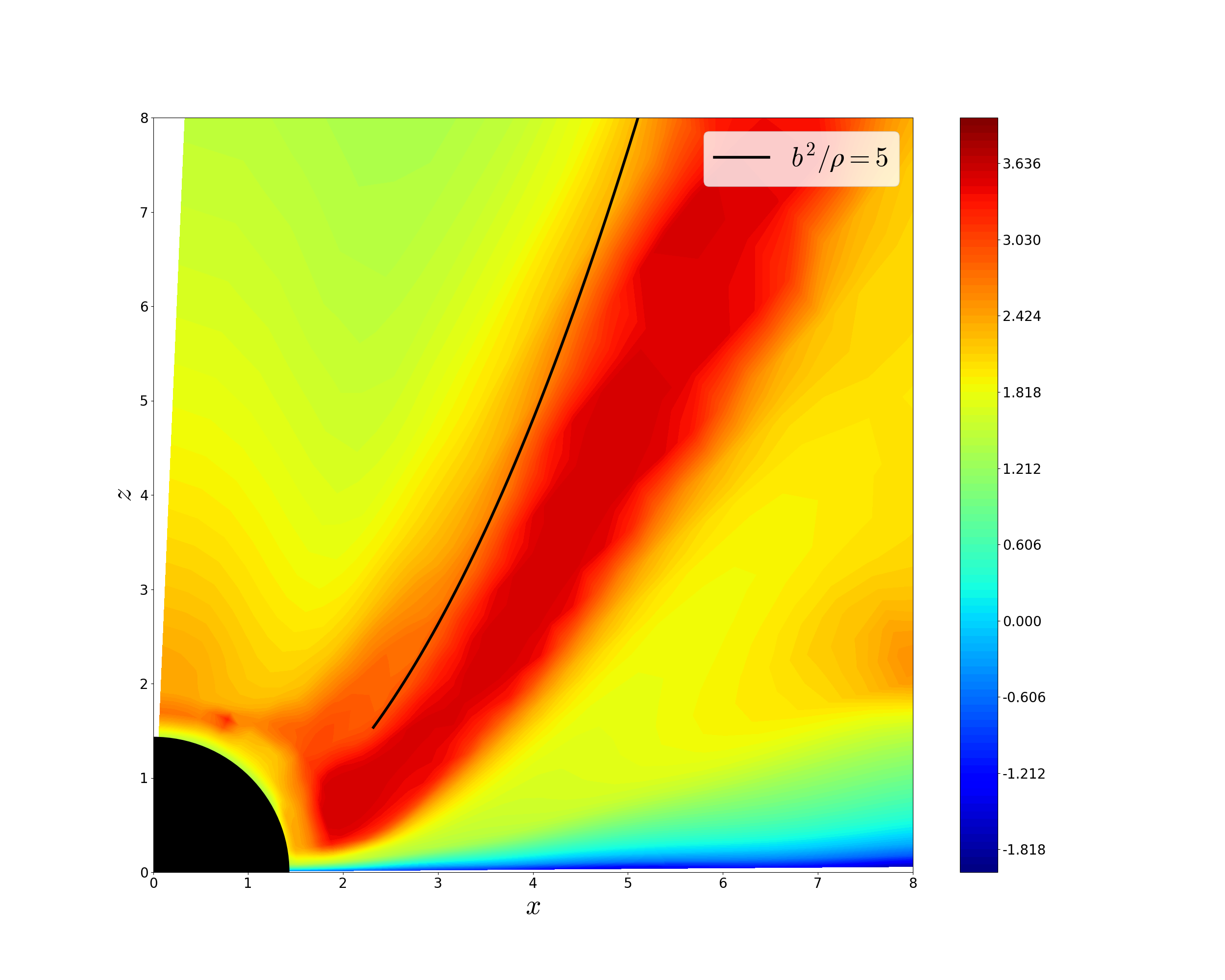}
    \includegraphics[width=0.45\textwidth]{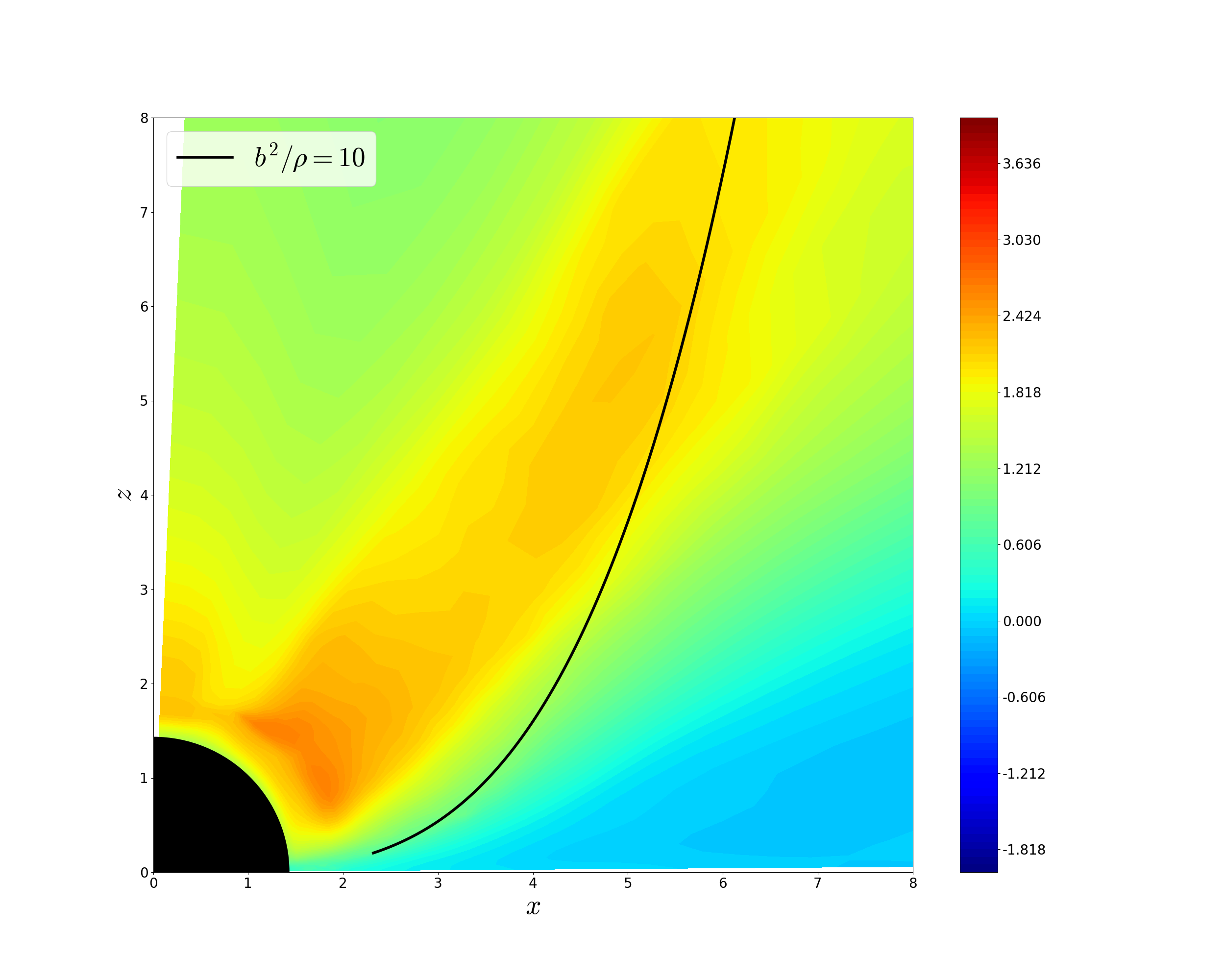}
    \caption{Distribution of $\log{\left(b^2/\rho\right)}$ (upper) and $\log{\left(b^2/p_g\right)}$ (lower) for 2000-3000$t_g$ (left) and 6000-7000$t_g$ averaged cases. The olive and magenta lines on the upper panels denote $b^2/\rho=10$ and $b^2/\rho=5$ respectively. The black line on the left-down panel denotes $b^2/\rho=5$ while the right-down one denotes $b^2/\rho=10$.}
    \label{fig:beta}
\end{figure}

In a typical BH accretion system, there are three primary regions: the disk, the corona, and the funnel region (which transforms into a jet in a strongly magnetized system). While GRMHD only provides the distributions of physical quantities throughout the space, the parameter $b^2/\rho$ is instrumental in
distinguishing these three regions. Generally, the disk is dominated by fluid part, hence $b^{2}/\rho\ll 1$,  while in the jet, the magnetic field is dominant, resulting in $b^{2}/\rho\gg 1$. A recommended critical value of $b^2/\rho$ on the jet surface is 5-10, as shown in the upper patterns of Fig.~\ref{fig:beta} \cite{HARM2004}. The lower patterns of Fig.~\ref{fig:beta} reveal that $b^2/p_g$ reaches its maximum near the jet surface (about $10^2$-$10^3$). This confirms that on the surface of the generated jet, the Alv{\'e}n velocity significantly exceeds the sound speed, resulting in a geometrically thin structure. The emission model discussed in section \ref{sec:emi_model} is applicative in this context. Practically, the thickness of the emission region ($\Delta\theta_{\rm jet}$) is determined by the $\theta$ displacement between where $b^2/\rho=5$ and $b^2/\rho=10$. The surface of the jet ($\theta_{\rm jet}(r)$) is chosen to be the one closest to the local maximum of $b^2/p_g$ within $b^2/\rho\simeq 5-10$.

\section{Numerical results}
\label{sec4}

\subsection{Evolution of Disk and Jet}
\label{sec:time_ave}

\begin{figure}
	\centering
	\includegraphics[width=250pt]{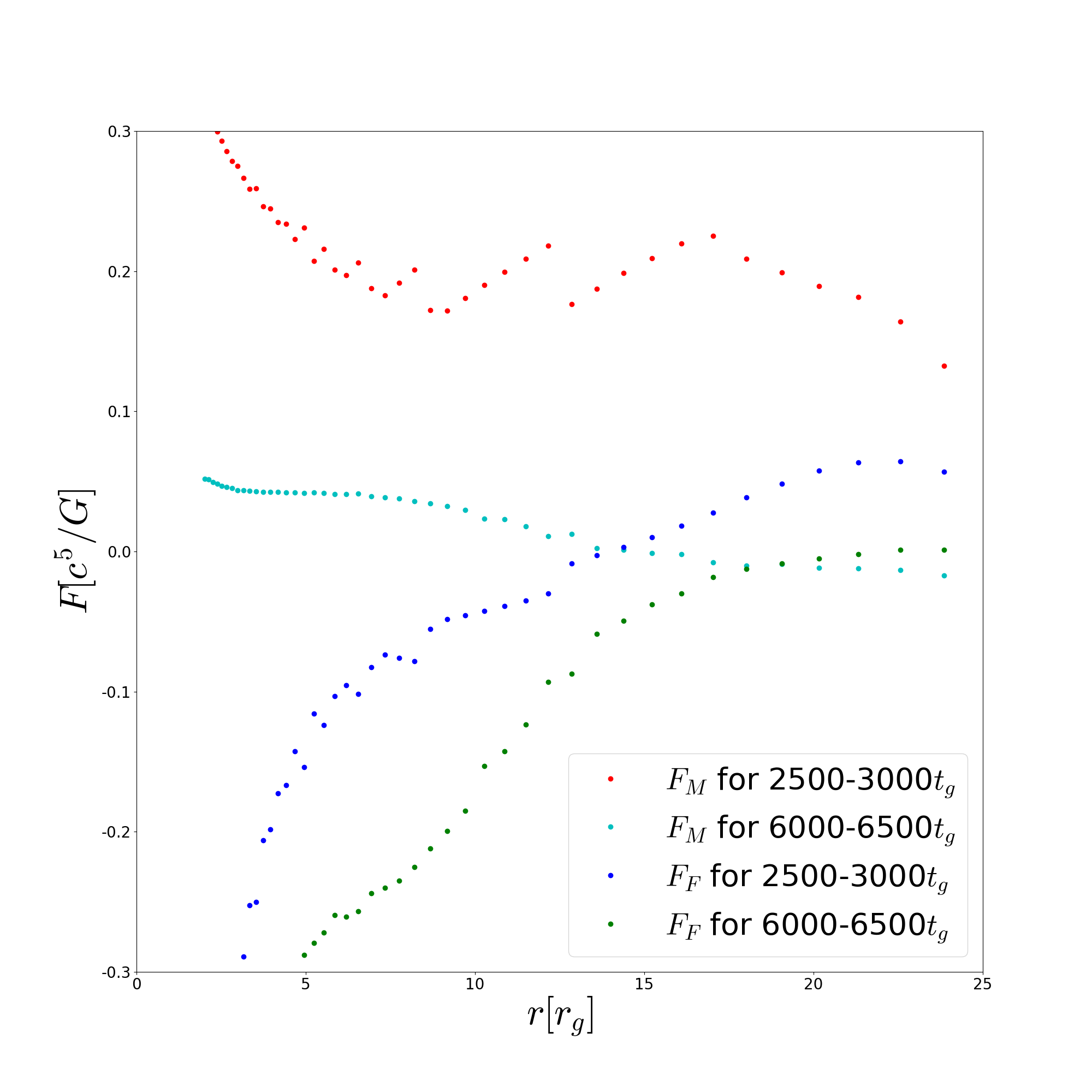}
	\caption{Time-averaged energy flux of the jet. $F_M$/$F_F$ stands for the flux of eletromagtic field/fluid respectively.}
	\label{fluxjet}
\end{figure}

It turns out that when the magnetic field in the accretion process is sufficiently large (as we initially set $p_{max}/b^{2}_{max}=2$), we can observe three distinct phases in the evolution, 
\begin{figure}
    \centering
    \includegraphics[width=350pt]{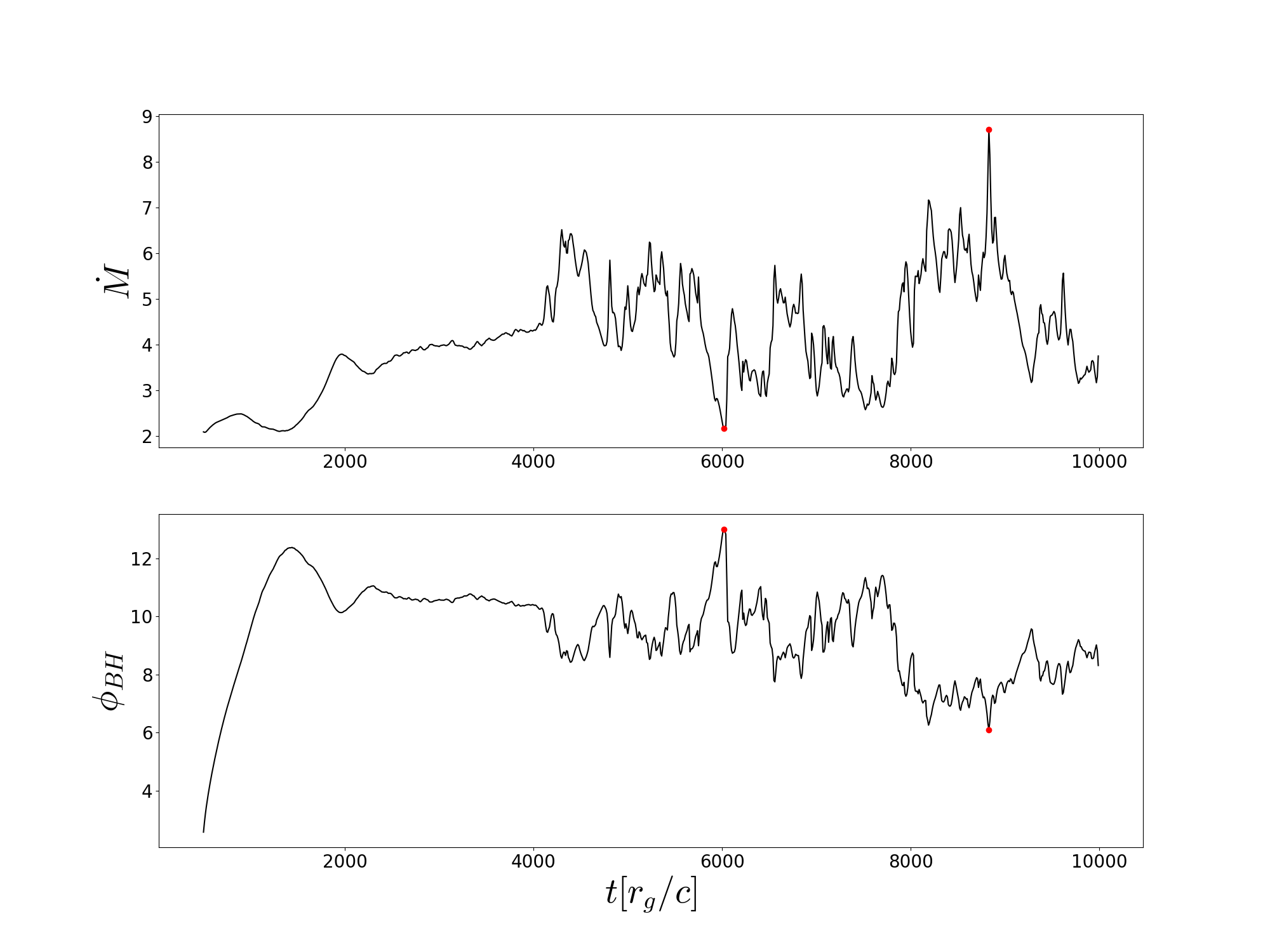}
    \caption{Evolution of the accretion rate (upper panel) and the dimensionless magnetic flux (lower panel) crossing the event horizon. The red dots denote 6020$t_g$ and 8830$t_g$, respectively.}
    \label{fig:mag_flux}
\end{figure}
as depicted in Fig.~\ref{fig:mag_flux}. Here, we demonstrate the evolution of the accretion rate and the dimensionless magnetic flux crossing the event horizon, which are defined as
\begin{equation}
    \label{rel_magflux}
    \begin{cases}
        \dot{M}=-\int_{\Omega_{\rm EH}} \sqrt{-g}\,\rho u^{r} d\theta\ d\phi
        \\
        \\
        \phi_{\rm BH}=\frac{1}{2\sqrt{\dot{M}}}\int_{\Omega_{\rm EH}} \sqrt{-g}\,|B^{r}| d\theta d\phi
    \end{cases}
\end{equation}
where $\Omega_{\rm EH}$ represents the surface of the event horizon. As shown in Fig.~\ref{fig:mag_flux}, given the parameters we have chosen, the first phase occurs at $t\lesssim 2000r_{g}/c$, during which both the accretion rate and the magnetic flux increase incrementally. 
\begin{figure}
    \centering	
    \begin{subfigure}[b]{\textwidth}
         \centering
         \includegraphics[width=0.25\textwidth]{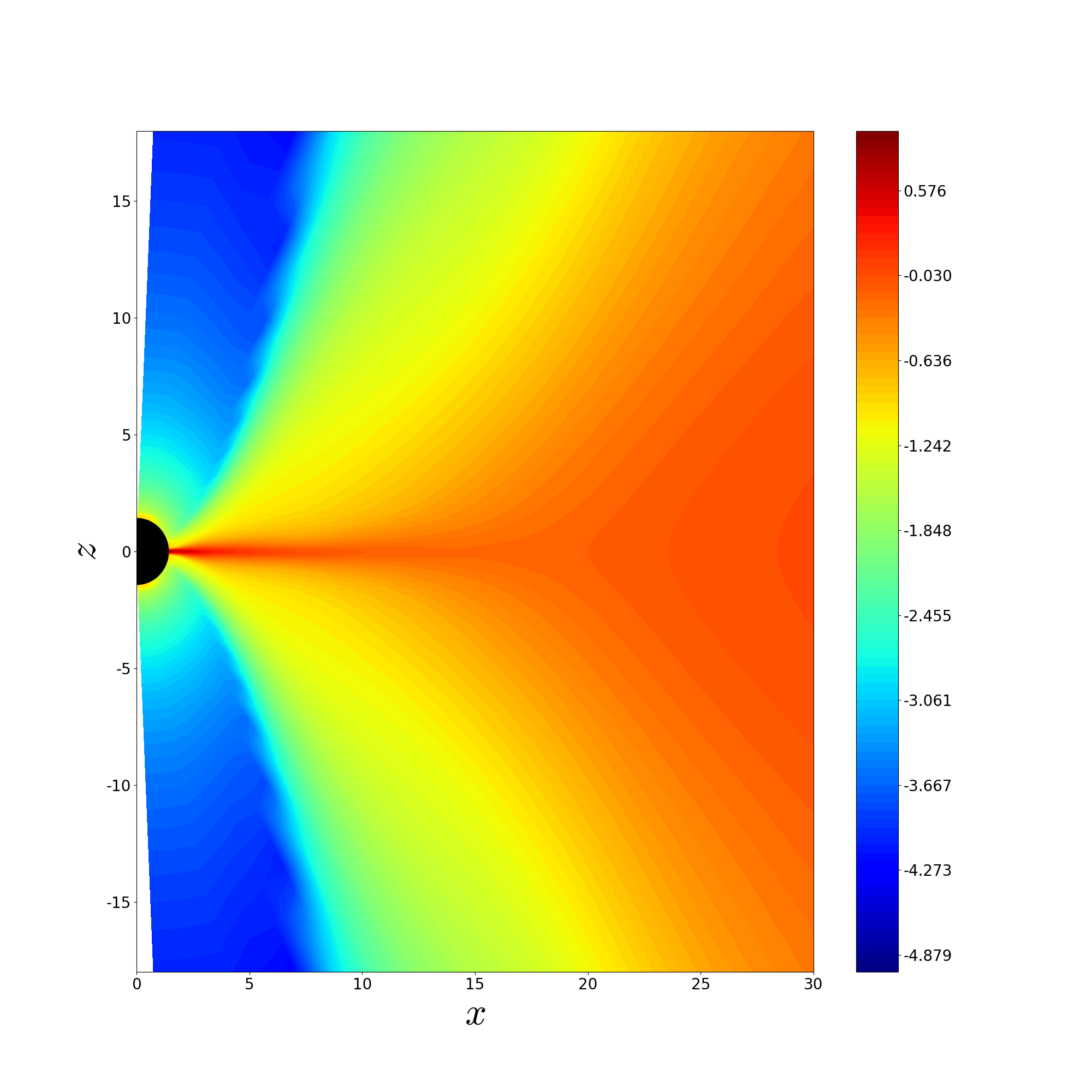}
         \includegraphics[width=0.25\textwidth]{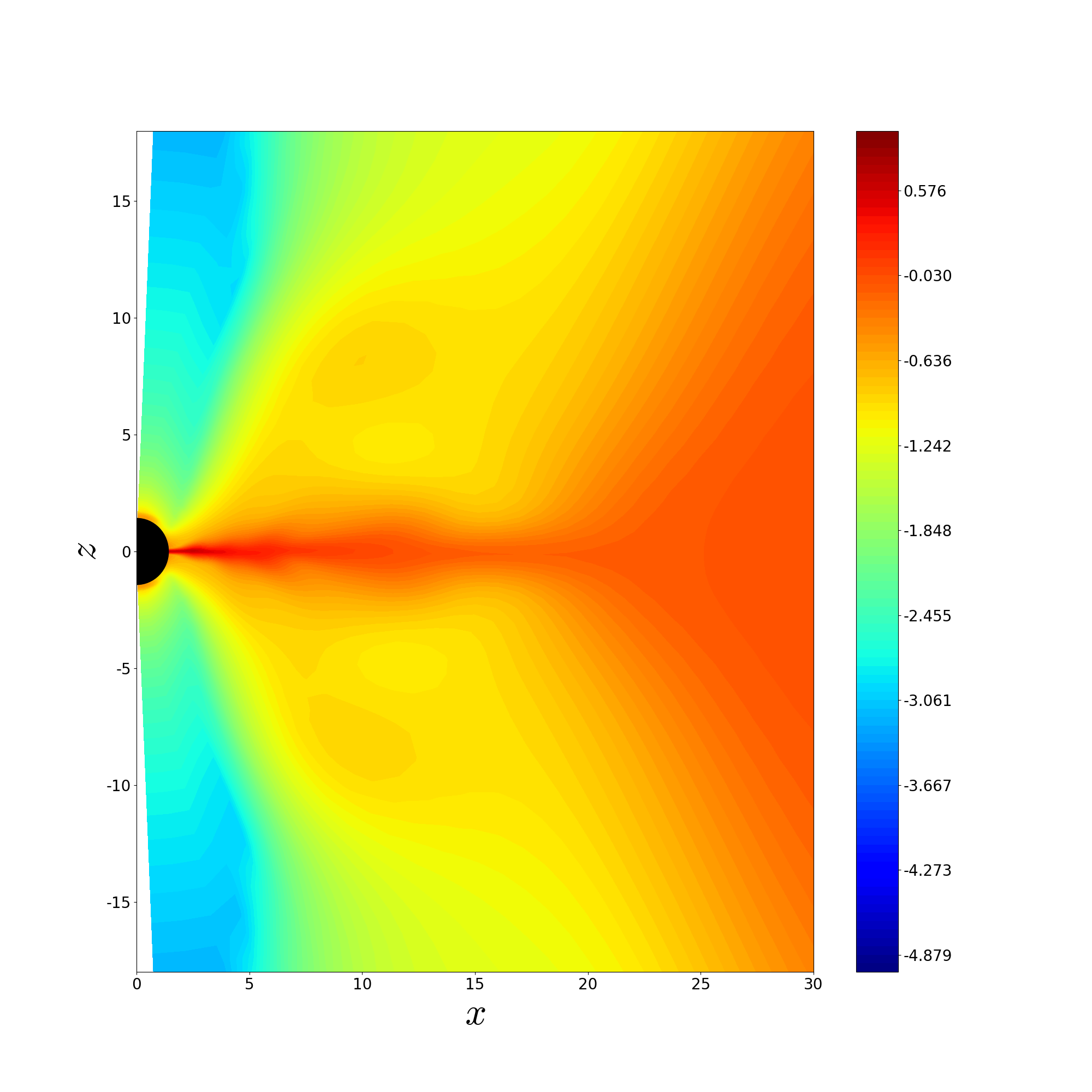}
         \includegraphics[width=0.25\textwidth]{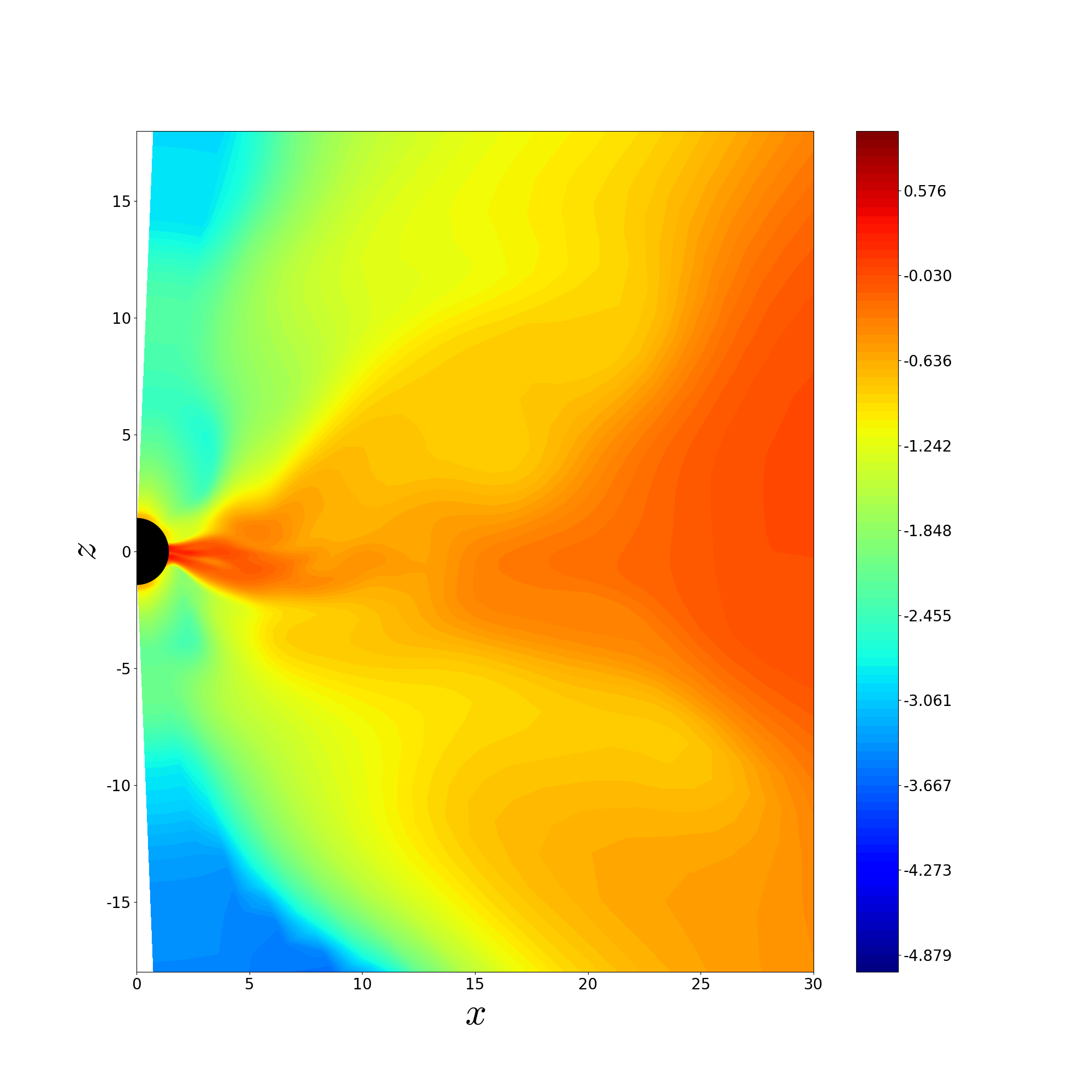}
         \caption{poloidal}
         \label{fig:rho_polo}
     \end{subfigure}
     \begin{subfigure}[b]{\textwidth}
         \centering
         \includegraphics[width=0.25\textwidth]{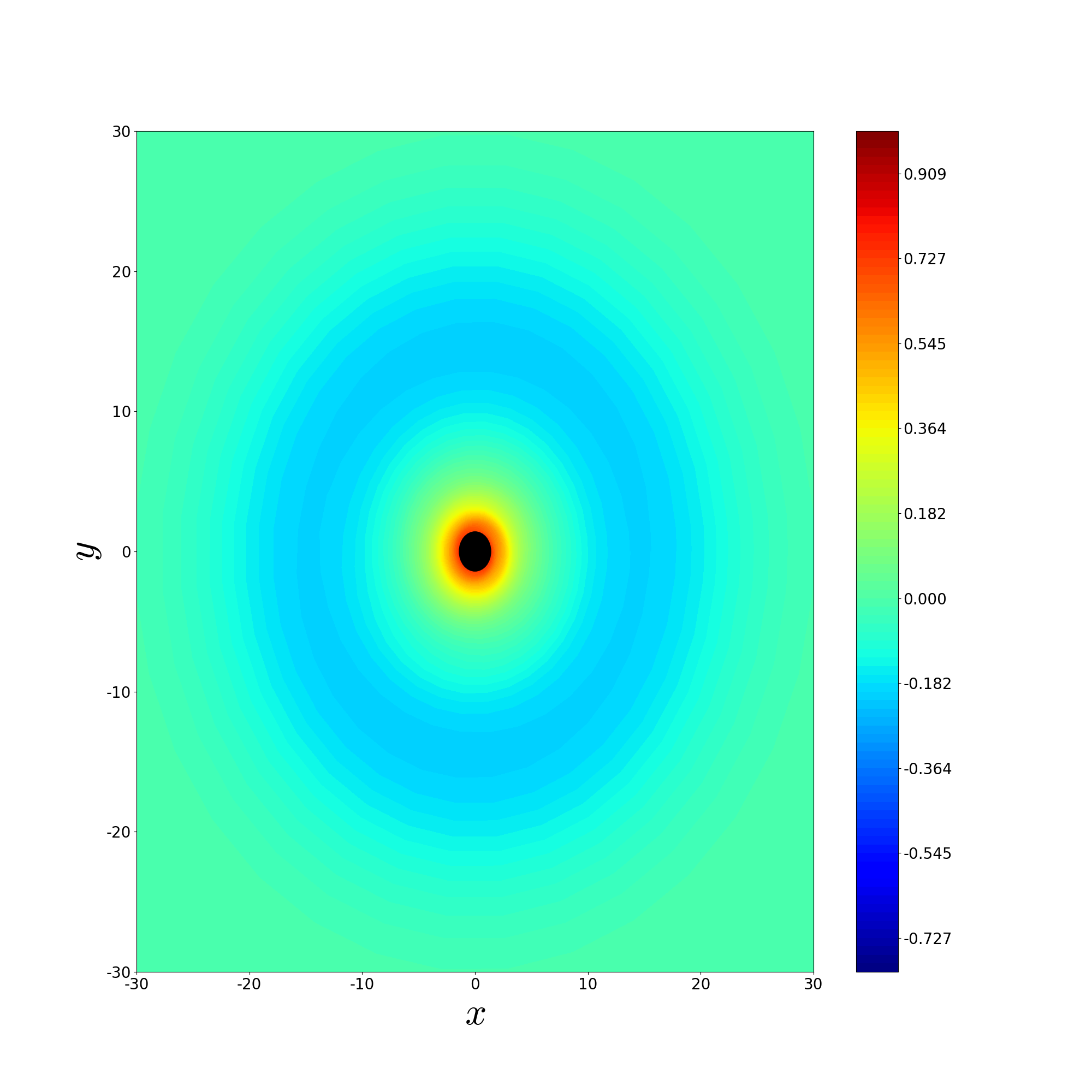}
         \includegraphics[width=0.25\textwidth]{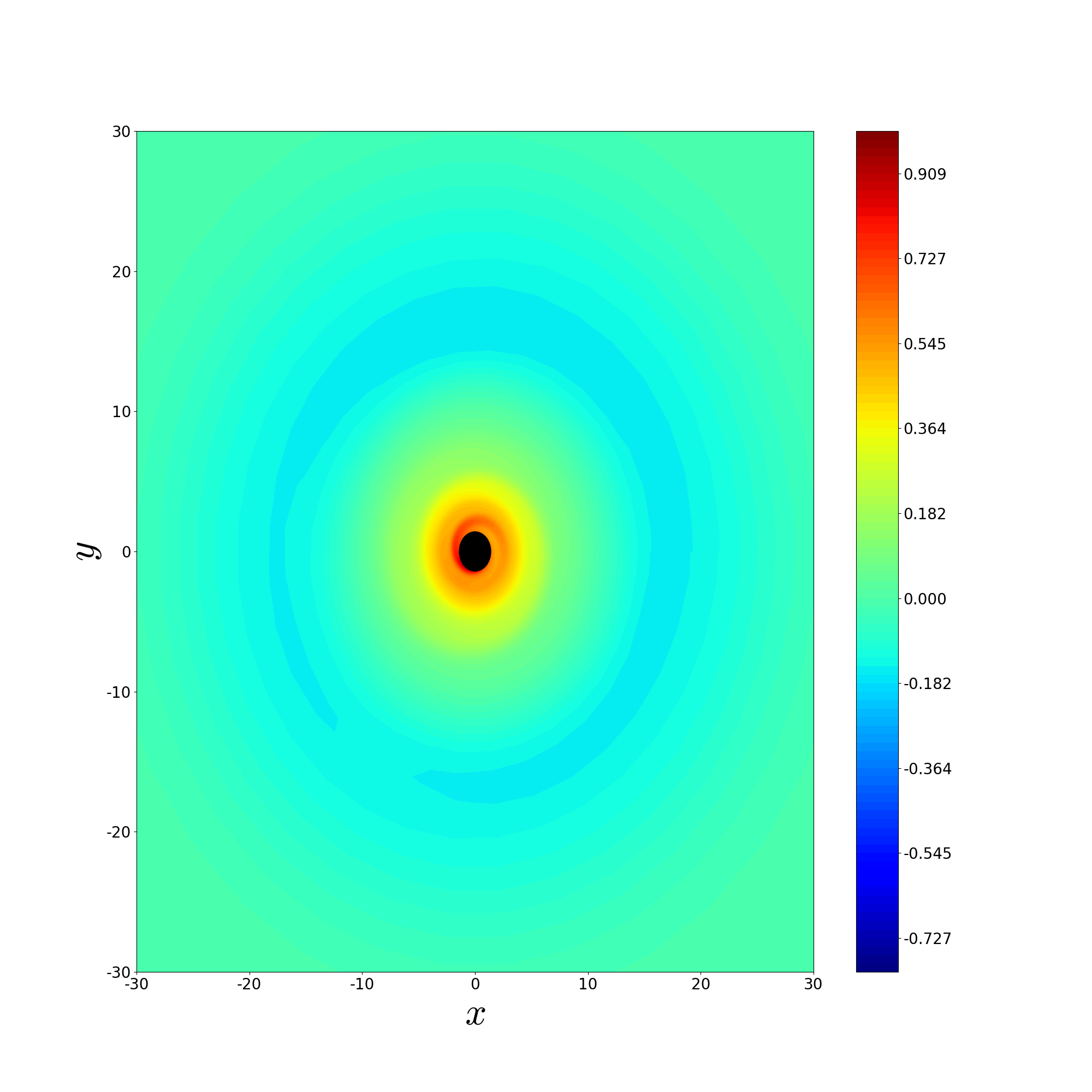}
         \includegraphics[width=0.25\textwidth]{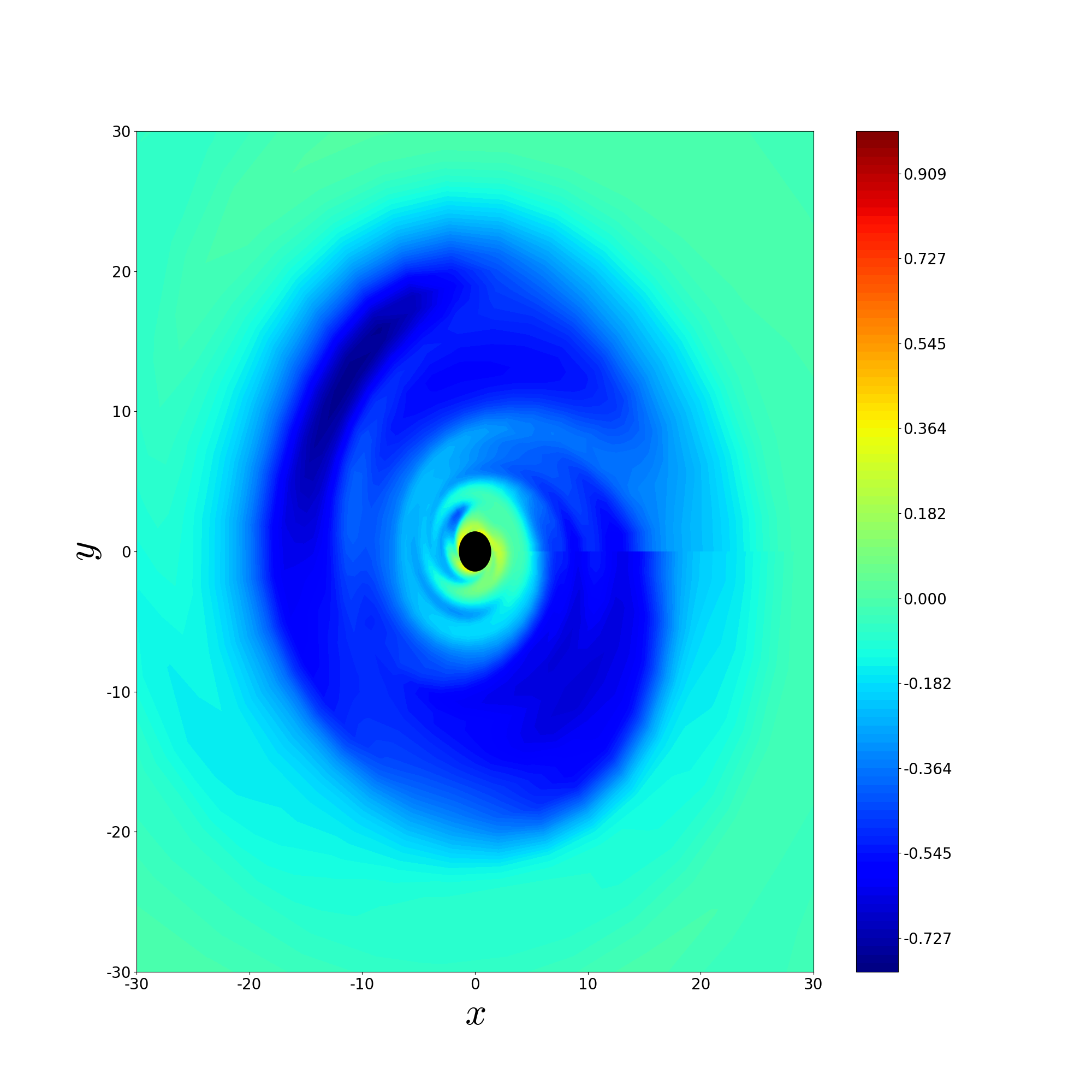}
         \caption{toroidal}
         \label{fig:rho_toro}
     \end{subfigure}
     \caption{2D mass-density (logarithmic) distribution of the accretion disk and the relativistic jet generated by GRMHD on poloidal and toroidal planes, for $t=1000t_g$(left), $t=3000t_g$(middle) and $t=8000t_g$(right).}
\end{figure}
During this phase, both the accretion flow and the magnetic field are weak near the event horizon (refer to the left panels in Fig.~\ref{fig:rho_polo} and Fig.~\ref{fig:rho_toro}). Beyond this period, when $2000r_{g}/c\lesssim t\lesssim 4000r_{g}/c$, the variations in the accretion rate and magnetic flux stabilize. Notably, the magnetic flux nearly reaches a saturation. Consequently, the fluid carrying the magnetic field encounters increased resistance to flow into the black hole due to the magnetic field's pushback \cite{SashaNote, ATchekho2011}. During this phase, a relativistic jet forms in the funnel region via the BZ mechanism. In Fig.~\ref{fluxjet}, we present the time-averaged energy flux of the jet in the near-horizon region (for both the saturation regime and the MAD regime). It is evident that the jet is tethered to the event horizon by magnetic field lines, hence the extraction of black hole rotational energy is dominated by electromagnetic processes, a characteristic feature of the BZ mechanism.
Within this regime, laminar flow predominates in the accretion process (refer to the middle panels in Fig.\ref{fig:rho_polo} and Fig.\ref{fig:rho_toro}). However, the presence of magneto-rotational instability (MRI) amplifies the magnetic field and disrupts the density distribution within the disk. As a result, turbulence forms and accumulates, leading to the creation of several regions of over-density or under-density.  

\begin{figure}
    \begin{subfigure}[b]{0.45\textwidth}
         \centering
         \includegraphics[width=\textwidth]{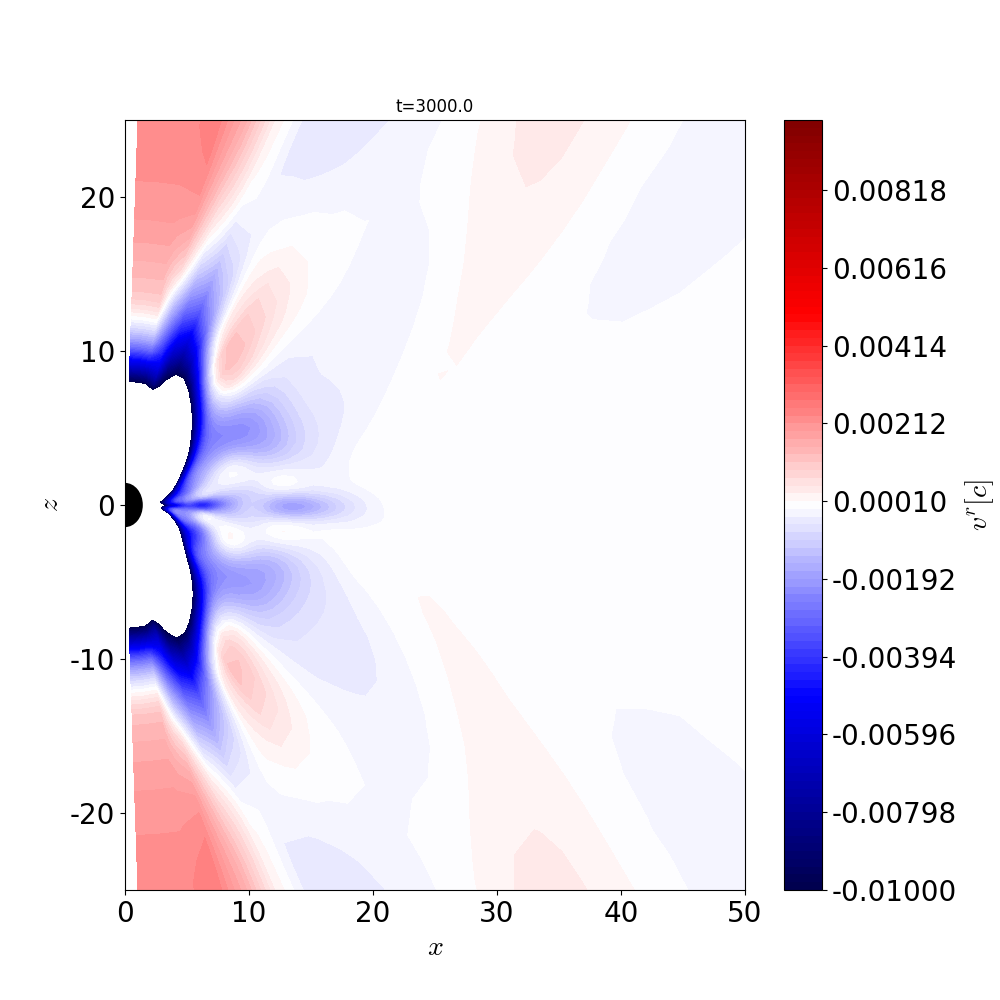}
         \caption{$t=3000$}
         \label{fig:vr300}
     \end{subfigure}
     \begin{subfigure}[b]{0.45\textwidth}
         \centering
         \includegraphics[width=\textwidth]{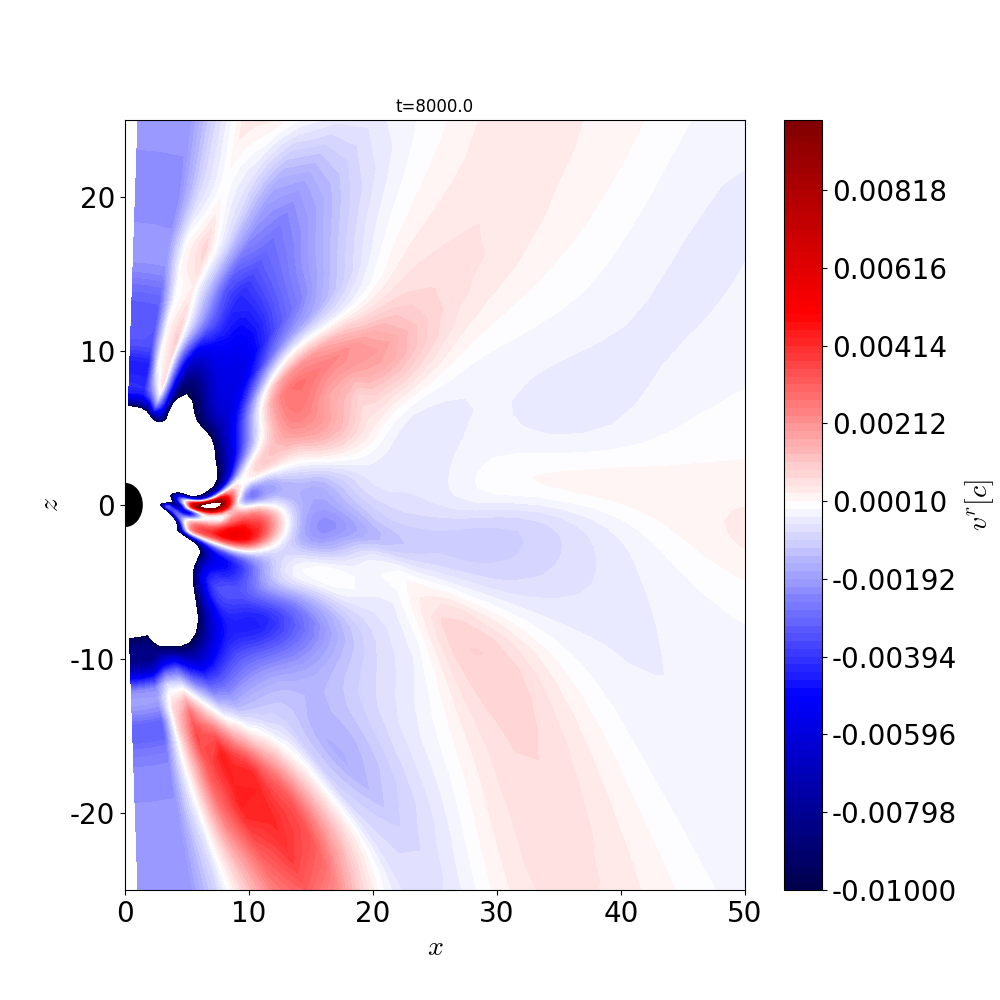}
         \caption{$t=8000$}
         \label{fig:vr800}
     \end{subfigure}
     \caption{2D radial velocity ($v_r$) distribution on poloidal planes at $3000t_g$ and $8000t_g$.}
\end{figure}

At $t\simeq 4000r_g/c$, it happens that gravity suddenly prevails over the magnetic field as the over-dense regions flow into the event horizon, or on the contrary, the magnetic field prevails over gravity as the under-dense regions flow into the event horizon. As a result, the accretion rate either increases or decreases, leading to an inflow of more or less fluid, respectively, carrying a magnetic field into the horizon. As we can see from Fig.~\ref{fig:mag_flux}, whenever the magnetic flux surpasses its saturation value, there follows a decrease of the accretion rate; while conversely  whenever the magnetic flux drops below its saturation value, there is an increase of accretion rate. Consequently, the accretion rate and the dimensionless magnetic flux oscillate chaotically around the saturation value. This dynamic, gravity and the magnetic field vying for dominance, is just like two children play on a seesaw. This phenomenon keeps the accretion system in a state of dynamic equilibrium, a state referred to as the MAD in the literature \cite{SashaNote, ATchekho2011}. Turbulence is predominant in the accretion disk near the horizon (as shown in the right panels of Fig.\ref{fig:rho_polo} and Fig.\ref{fig:rho_toro}), and outflows form in both the jet and the disk (as illustrated in Fig.~\ref{fig:vr800}).

\subsection{Jet Images}
\label{sec:result}
We continue our study by separately examining the images of the jet surface in both the saturation and MAD regimes. We average all physical values for $t$ from 2000$t_g$ to 4000$t_g$ and from 4000$t_g$ to 10000$t_g$.  For M87*, which has a mass of about $5\times10^9 M_{\odot}$, $10^3t_g$ corresponds to roughly $7hr$. We also consider two specific moments corresponding to the lowest and the highest accretion rates within $10000 r_g/c$, as shown in Fig. \ref{fig:mag_flux}. 

It is worth noting that the magnitude of physical quantities in GRMHD simulation is far away from the reality. Thus, before taking imaging by using GR ray-tracing, we rescale all physical quantities so that the accretion rate aligns with the recommended accretion rate of M87*: $\dot{M}\sim 10^{-3}M_{\odot}\,yr^{-1}$ (see appendix \ref{sec:rescale} for details). As previously discussed, the jet emission is primarily concentrated on its surface, which approximates a two-dimensional structure. Thus, it becomes considerably easier to fit the numerical data generated by GRMHD with a smooth function (we choose a power function of the form $f=Cr^A$) when we plot using GR ray-tracing. The method of least squares is applied here. Technically, the behavior of jet (as well as accretion flow) extremely close to the event horizon would be complicated, making it unsuitable to describe all the regions in an uniform way. It propels us to cut our curve fitting on a radius. As an empirical choice, we cut the curve fitting on the radius of the ISCO. For more details about the fitting curves, please refer to appendix \ref{sec:result_curve_fit}.
\begin{figure}
         \centering
         \includegraphics[width=\textwidth]{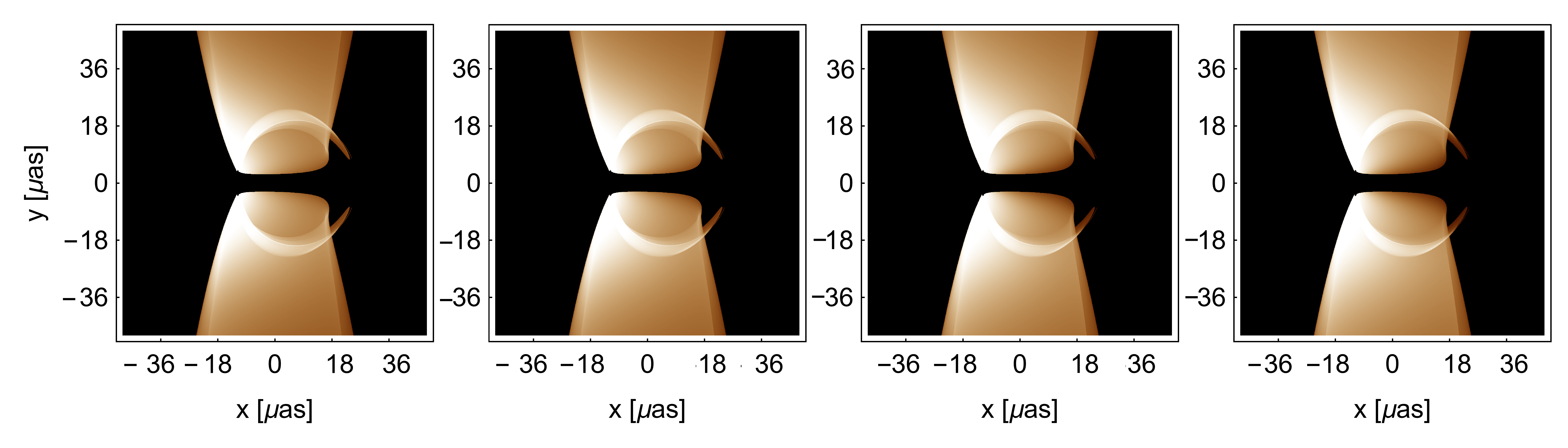}
         \includegraphics[width=\textwidth]{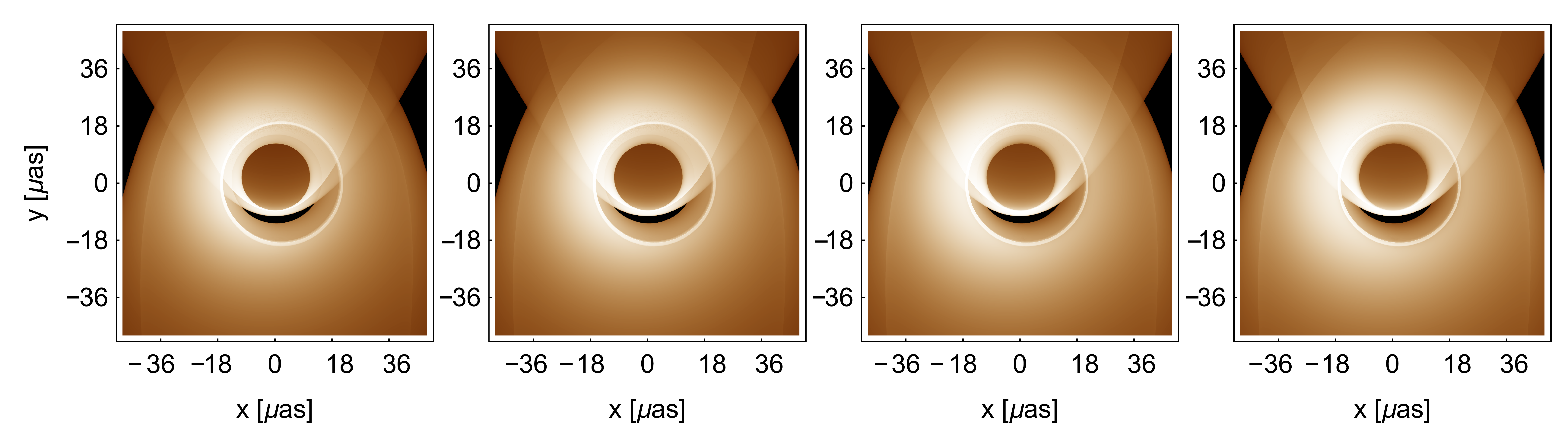}
         \includegraphics[width=\textwidth]{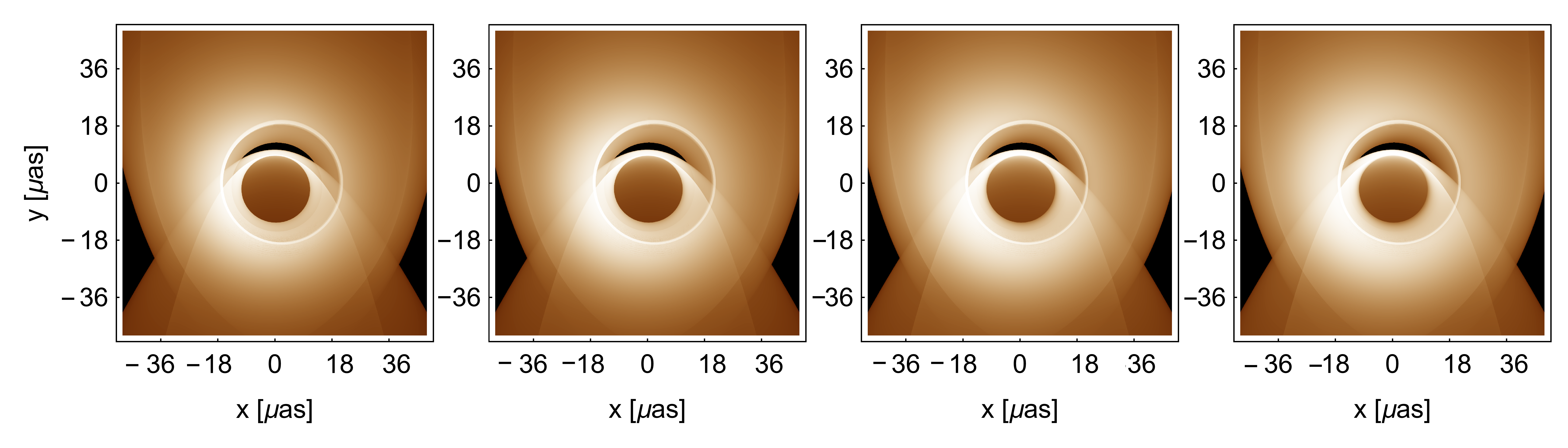}
         \caption{Images of the time-averaged jet within the saturation regime for $\t_{\rm obs}=90^{\circ}$ (upper), $\t_{\rm obs}=17^{\circ}$ (middle) and $\t_{\rm obs}=163^{\circ}$ (lower). From left to right columns, the observational frequency is $\nu=43\,,86\,,230\,,345\,\mm{GHz}$.}
         \label{Lam}
\end{figure}
\begin{figure}
	   \centering
	   \includegraphics[width=0.4\textwidth]{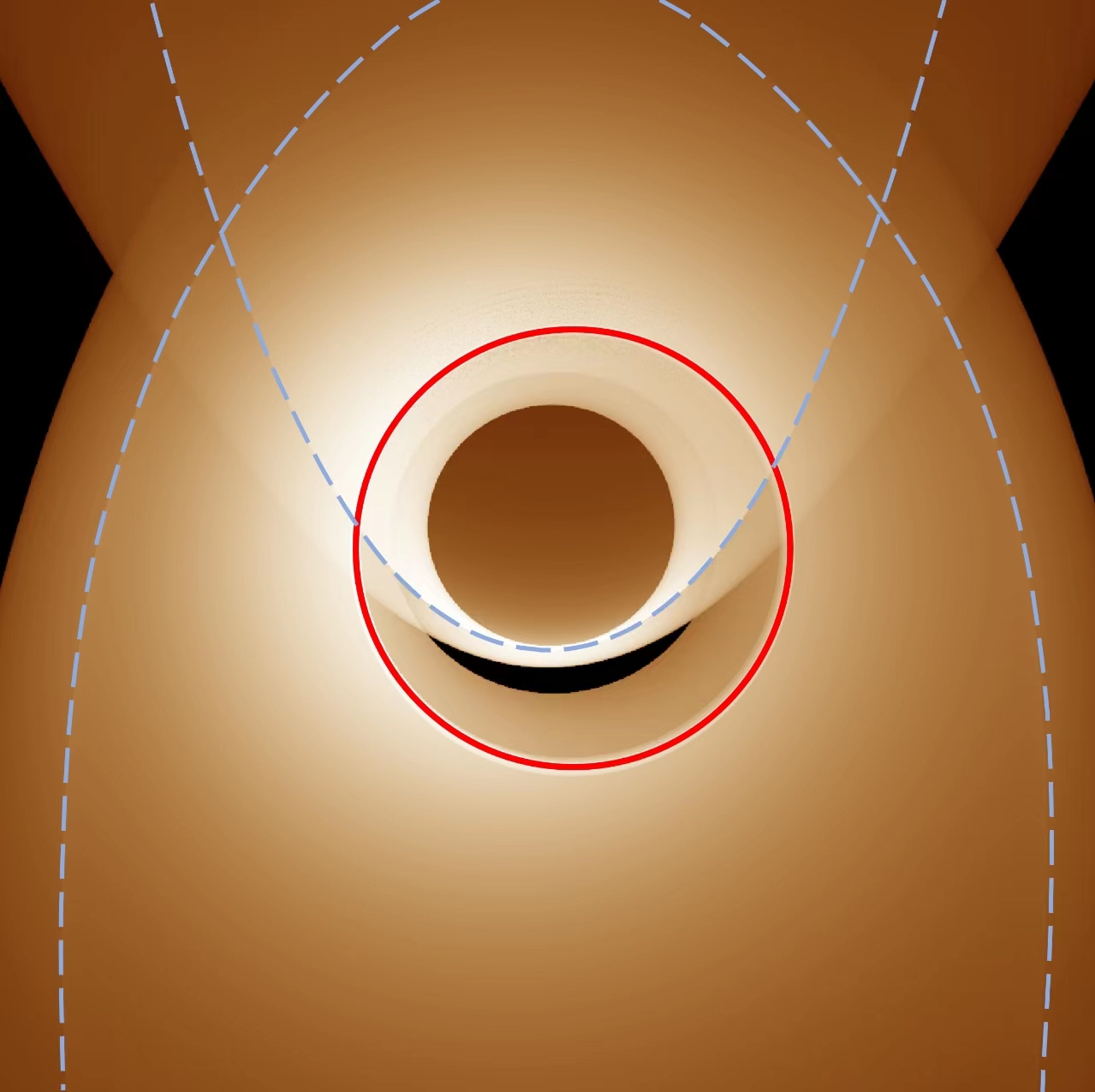}
        \caption{The strokes of photon ring (red solid circle) and brighter U-shaped lines (blue dashed lines).}
        \label{fig:Uline}
\end{figure}
\begin{figure}
\centering
         \includegraphics[width=\textwidth]{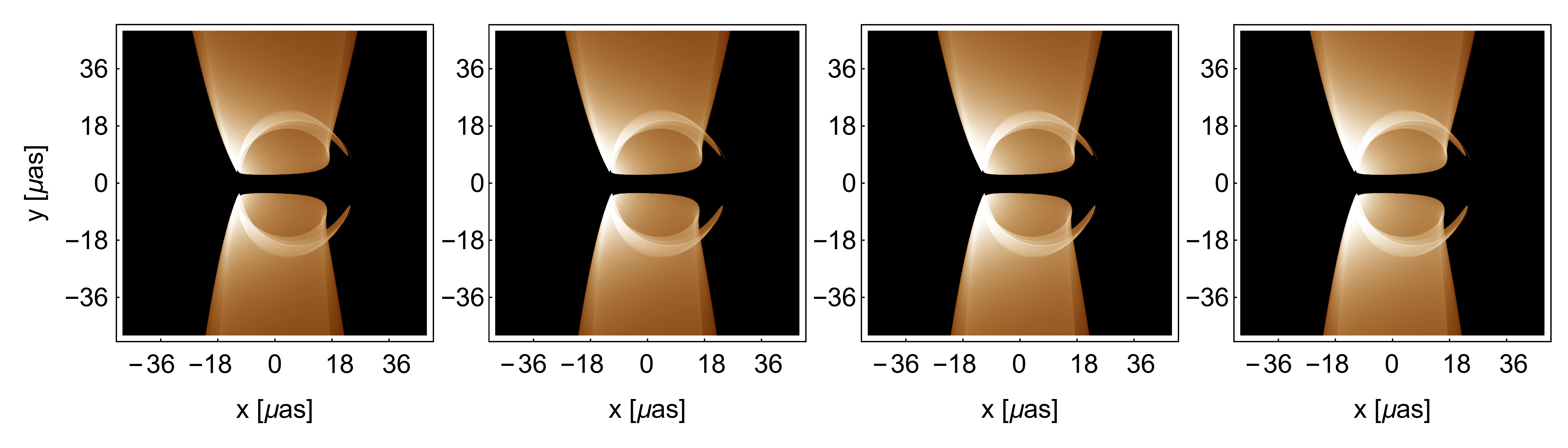}
         \includegraphics[width=\textwidth]{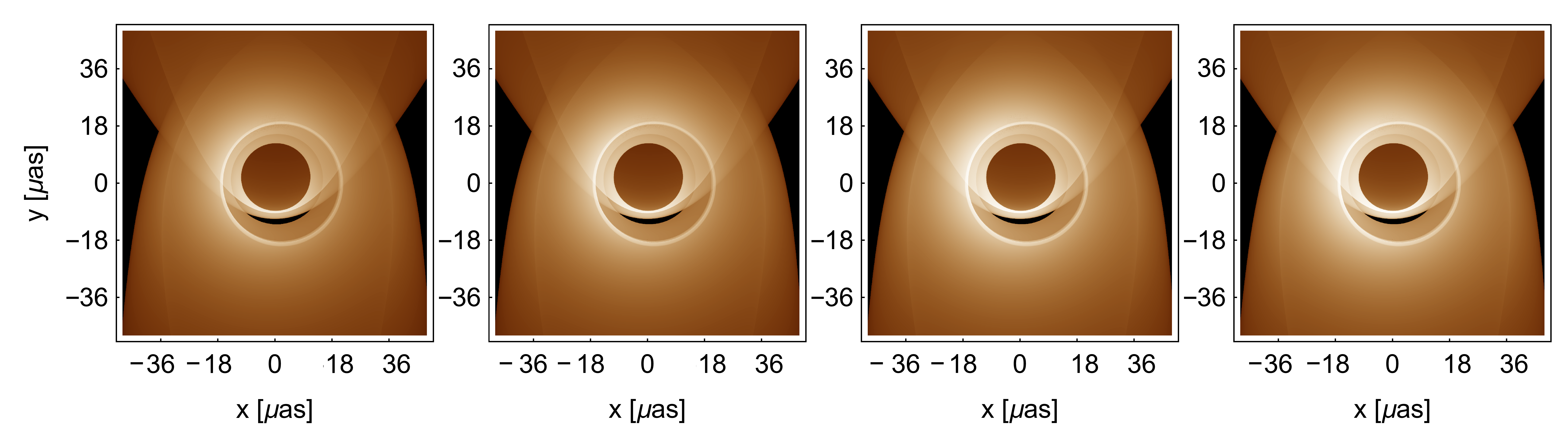}
         \includegraphics[width=\textwidth]{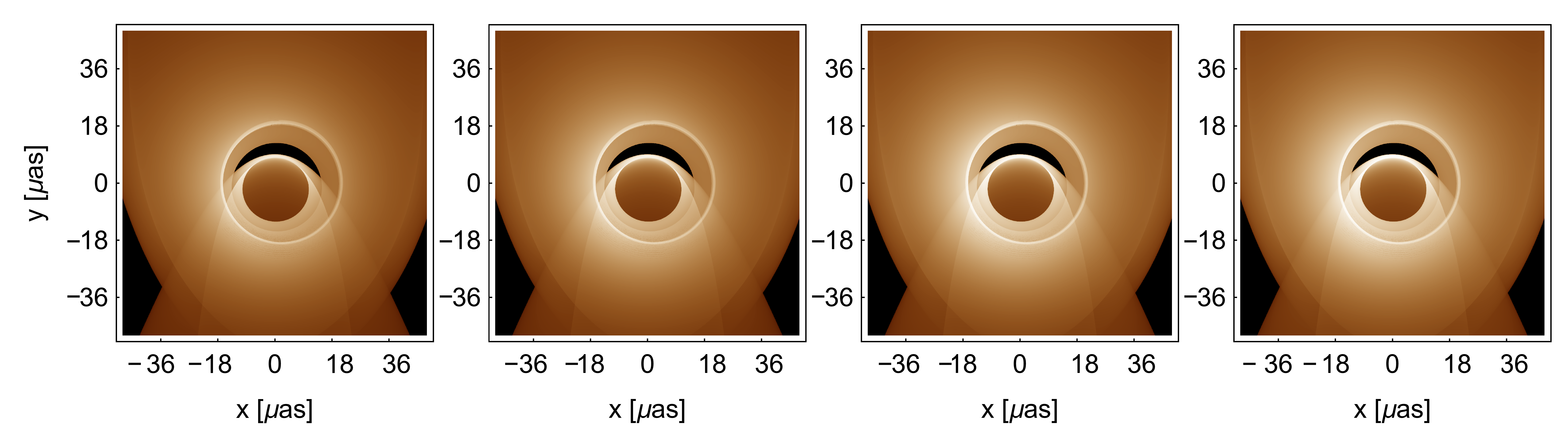}
     \caption{Images of the time-averaged jet in the MAD regime for $\t_{\rm obs}=90^{\circ}$ (upper), $\t_{\rm obs}=17^{\circ}$ (middle) and $\t_{\rm obs}=163^{\circ}$ (lower). From left to right columns, the observational frequency is $\nu=43\,,86\,,230\,,345\,\mm{GHz}$.}
     \label{Tur}
\end{figure}

In Fig.~\ref{Lam}, we present the images of the time-averaged jet within the saturation regime, observed at angles of $90^{\circ}$, $17^{\circ}$ and $163^{\circ}$, respectively.  The primary images at $\theta_{\rm obs}=90^{\circ}$ depict the shape of the jet surface in proximity to the event horizon. The significantly brighter lines on the left sides, due to the beaming effect, indicate the direction of the BH spin. Furthermore, these images demonstrate intuitively that the two branches of the jet are nearly symmetrical.

The images observed from $17^{\circ}$ and $163^{\circ}$ are particularly intriguing (see also Fig.~\ref{Tur}), as we can obviously see two brighter U-shaped lines on two branches of jet on each plot. As shown in Fig.~\ref{fig:Uline}, we zoom in on the U-shaped lines and the photon ring of middle-left panel of Fig.~\ref{Lam}, as an example. It is crucial to emphasize that these U-shaped lines are not a result of the beaming effect. Instead, they originate from the light rays whose paths closely graze the jet surface and traverse a considerably longer distance within the emission region. This allows them to accumulate a higher intensity from synchrotron radiation compared to the light rays that merely pass through the jet surface. This is a unique characteristic of geometrically and optically thin jets where the surface is the sole emission source. One can expect that a thinner surface produces sharper, brighter lines. However, these lines will vanish if absorption becomes substantial or if the entire funnel region turns into an emission source.

In Fig.~\ref{Tur}, we present the images of the time-averaged jet within the MAD regime, viewed from angles of $90^{\circ}$, $17^{\circ}$ and $163^{\circ}$ respectively. As previously mentioned, the entire accretion system exhibits significant oscillations in this scenario. However, by applying a time average, we observe that the jet images exhibit only minor differences from those in the saturation regime. It is easy to comprehend that oscillations will be largely eliminated in time-averaged images if the accretion system maintains a state of dynamical equilibrium.

\begin{figure}
         \centering
         \includegraphics[width=\textwidth]{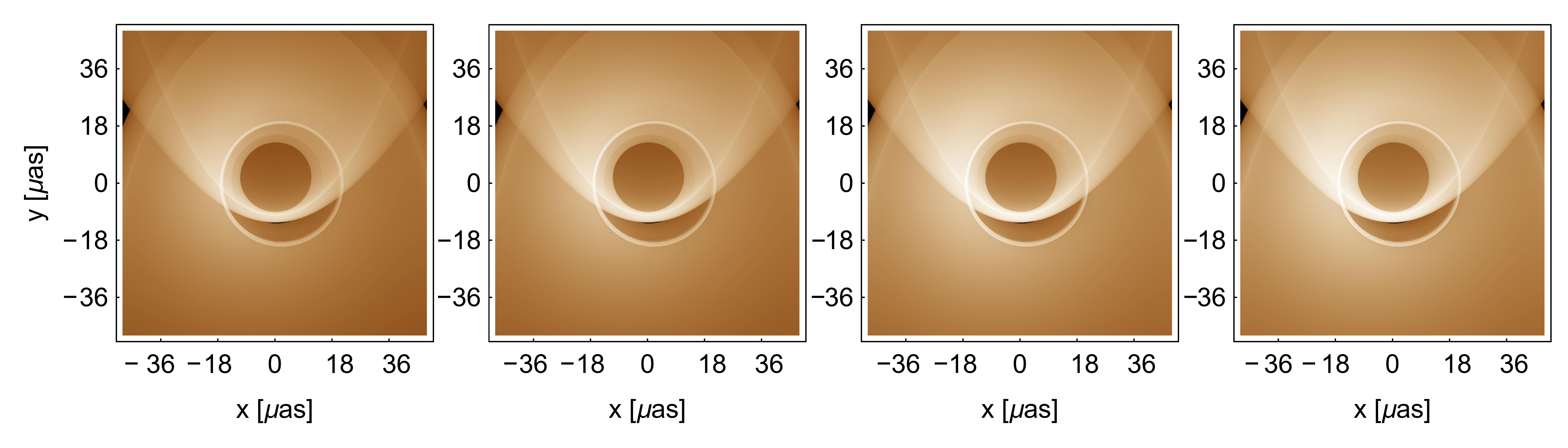}
         \includegraphics[width=\textwidth]{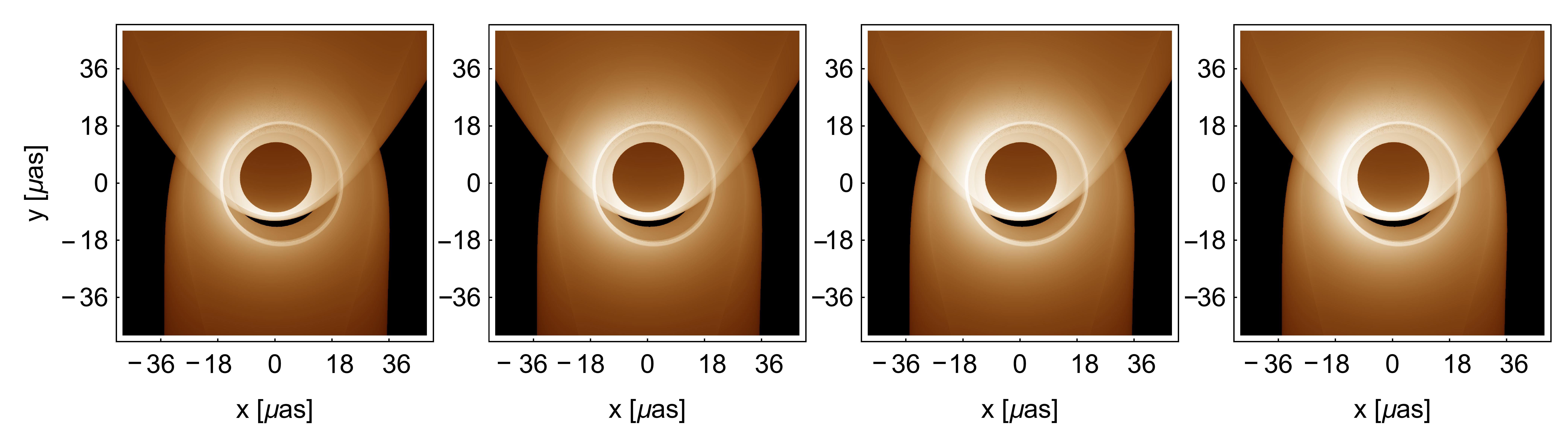}
         \includegraphics[width=\textwidth]{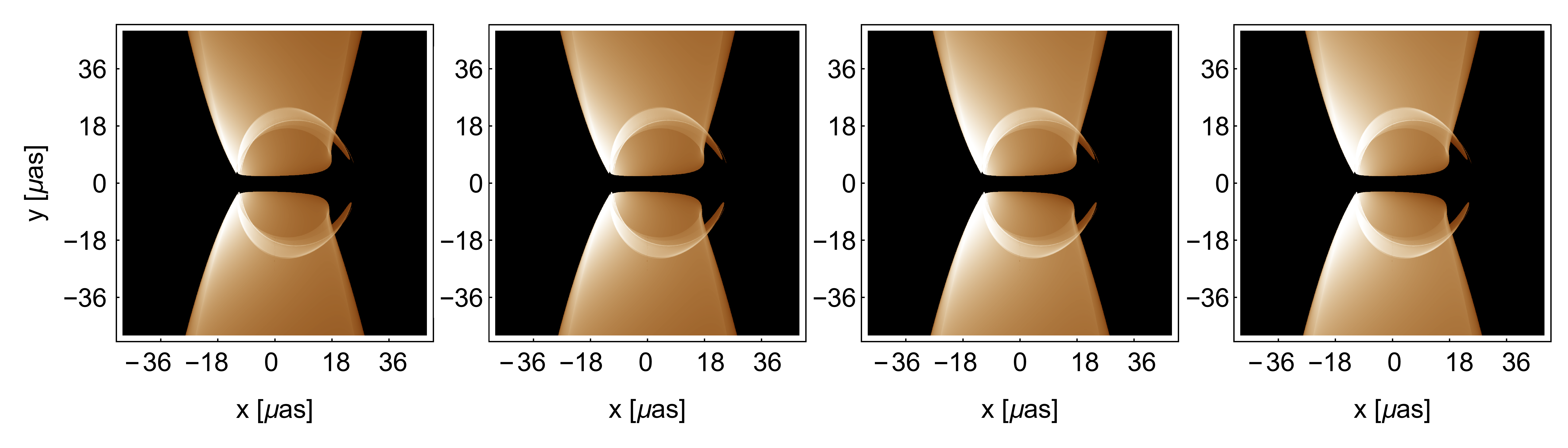}
         \includegraphics[width=\textwidth]{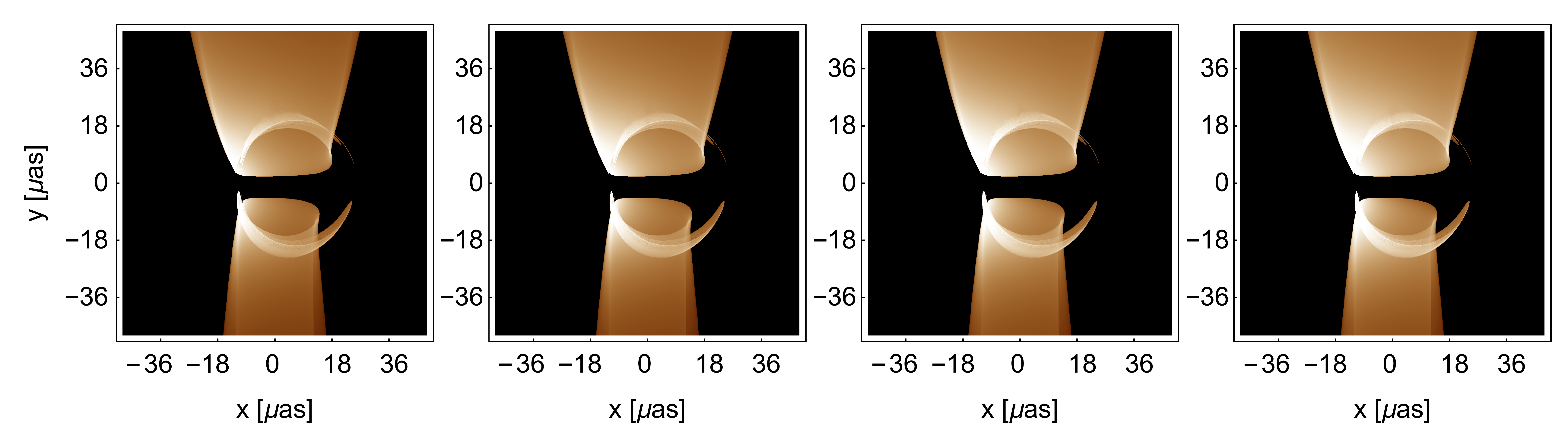}
     \caption{Snapshots of the jet within MAD regime with observational angle $\t_{\rm obs}=17^{\circ}$ for $t=6020 r_g/c$ (first row) and $t=8830 r_g/c$ (second row) and with $\t_{\rm obs}=90^{\circ}$ for $t=6020 r_g/c$ (third row) and $t=8830 r_g/c$ (fourth row). From left to right columns, the observational frequency is $\nu=43\,,86\,,230\,,345\,\mm{GHz}$.}
         \label{snap}
\end{figure}

Snapshots from two distinct moments in the MAD regime are displayed in Fig~\ref{snap}. Contrary to the time-averaged results, a clear asymmetry between the two branches is evident, which is a consequence of chaotic oscillation. Other features, such as the U-shaped brighter lines, the beaming effect, and the photon ring, are similar to those observed in the time-averaged results. 

\begin{figure}
\centering
	\begin{tikzpicture}
	\node at (0,0)
	{\includegraphics[width=\textwidth]{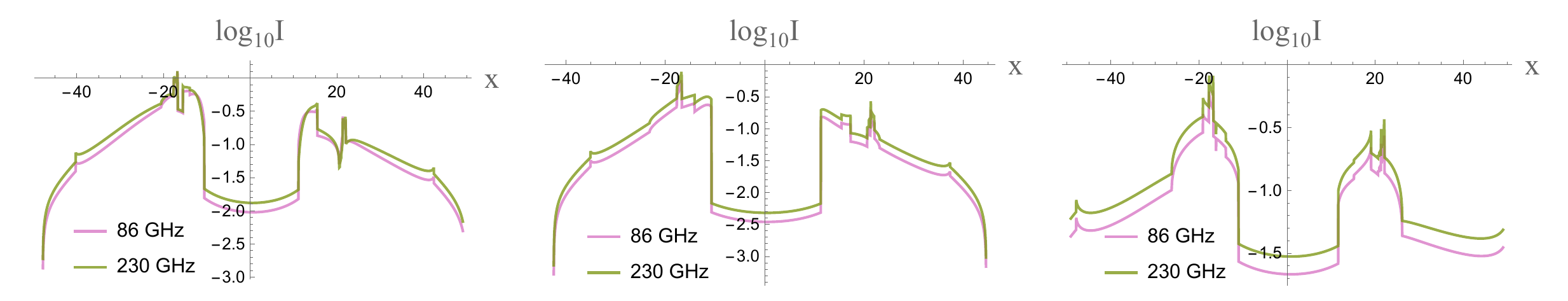}};
	\node[below left] at (-4.5,-0.1)
	{\LARGE\color{blue}$\uparrow$};
	\node[below left] at (-4.8,1.2)
	{\LARGE\color{red}$\downarrow$};
	\node[below left] at (-3.5,0.5)
	{\LARGE\color{red}$\downarrow$};
	\end{tikzpicture}
	\includegraphics[width=\textwidth]{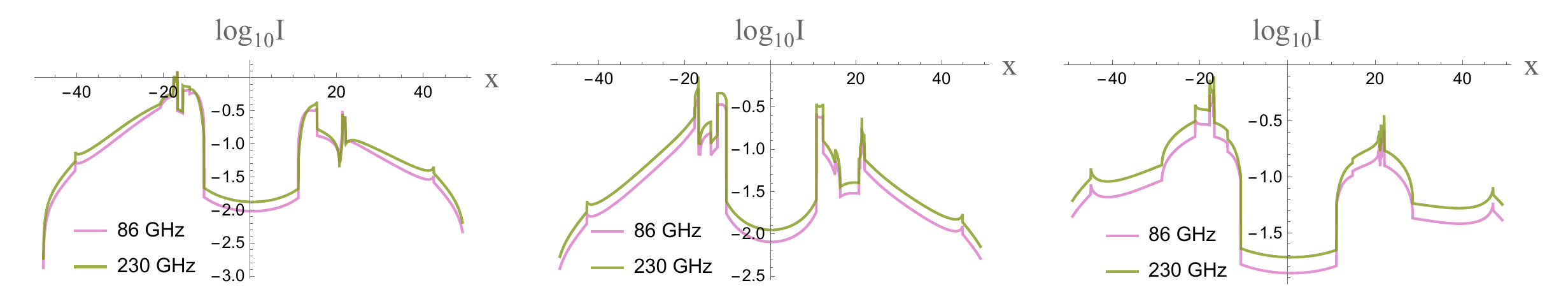}
	\caption{Intensity cuts on $y=0$ for $\t_{\rm obs}=17^{\circ}$ (top) and $\t_{\rm obs}=163^{\circ}$ (bottom). The intensity is normalized by $10^{-35}$. Left panels: time-averaged results of saturation regime. Middle panels: time-averaged results of MAD regime. Right panels: snapshot of MAD regime at $t=6020r_g/c$.}
	\label{intensitycut}
\end{figure}

In Fig. \ref{intensitycut}, we further present the intensity cut along the $x$-axis ($y=0$) for time averages within both the saturation (left panels) and MAD regimes (middle panels) and the snapshot at 6020$t_g$ (right panels). For the time-averaged results, it is evident that the two regimes exhibit only minor differences. However, for a snapshot in the MAD regime, the distribution of observed intensity may vary significantly, as depicted in the right panel of Fig.~\ref{intensitycut}. Generally, photon rings are represented as peaks near $x=20[\mu as]$ (see blue upward arrow on upper left panel of Fig.~\ref{intensitycut}) while U-shaped lines on the two branches manifest as smaller peaks near $20[\mu as]$ and $40[\mu as]$ (see red downward arrows on upper left panel of Fig.~\ref{intensitycut}). Besides, the intensity on the positive $x$-axis is higher, a result of the beaming effect.

\begin{figure}
	\centering
	\begin{subfigure}[b]{0.45\textwidth}
         \centering
         \includegraphics[width=\textwidth]{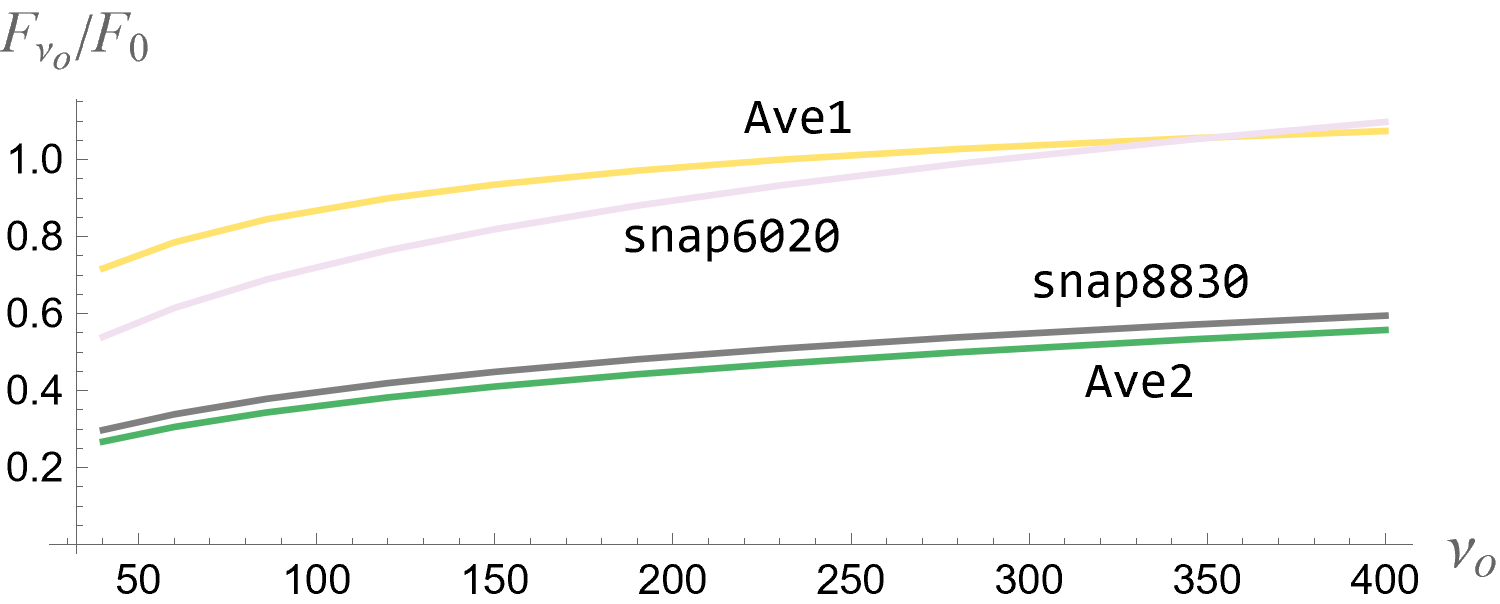}
         \caption{$\theta_{\rm obs}=17$}
         \label{fig:flux17}
     \end{subfigure}
     \begin{subfigure}[b]{0.45\textwidth}
         \centering
         \includegraphics[width=\textwidth]{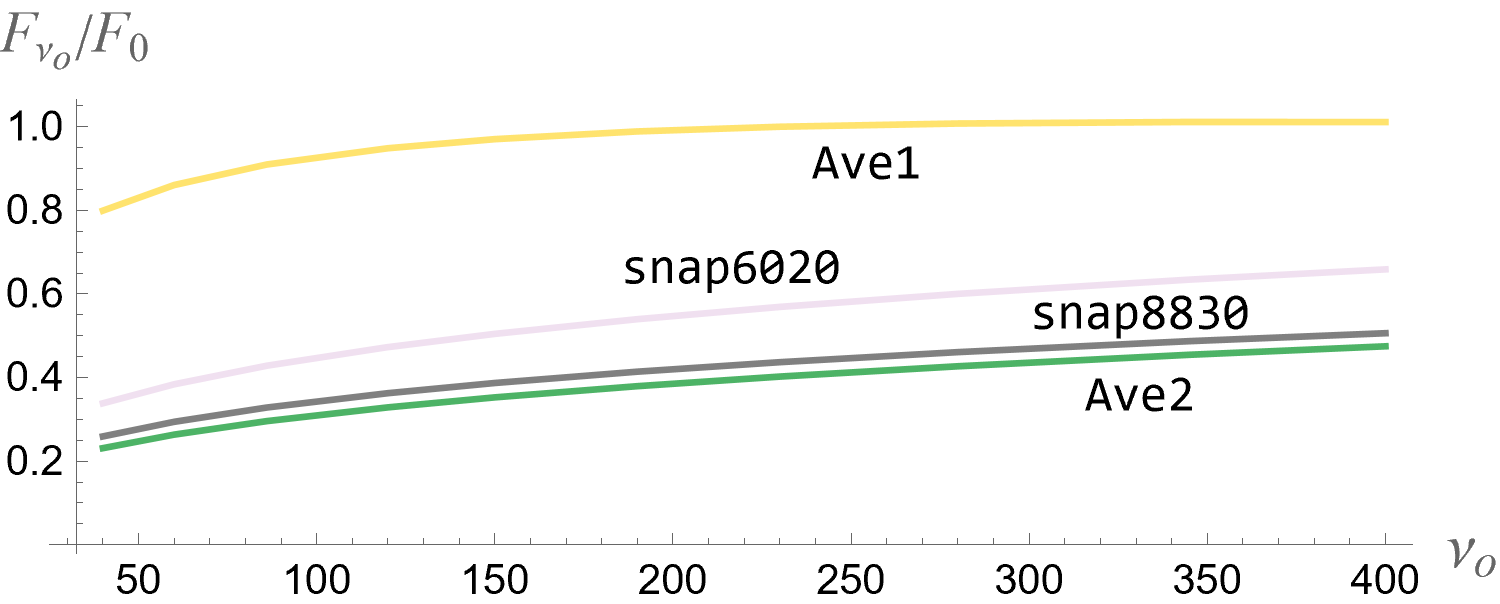}
         \caption{$\theta_{\rm obs}=90$}
         \label{fig:flux90}
     \end{subfigure}
	\caption{Observed total flux for different regimes and snapshots, normalized by the flux of saturation regime at $230$ GHz. Here the results of time-averaged jet of the saturation regime and MAD regime are denoted by Ave1 and Ave2 respectively.}
     \label{fig:tflux}
\end{figure}

In addition, we observe that all images are largely insensitive to the observational frequency. We also plot the variation of the observed total flux with the frequency in Fig.~\ref{fig:tflux}, which is given by integrating the intensity on the observer's screen, $F_{\n} = \int I_{\n}dxdy$. Fig.~\ref{fig:tflux} reveals that although the total flux increases with higher frequencies, the growth rate is quite small. This is primarily due to the absence of absorption. As can be seen from Fig.~\ref{fig:tflux}, the emission from time-averaged jet within the saturation regime (denoted as Ave1) is significantly higher than that in the MAD regime (denoted as Ave2), and the total flux at $t=6020t_g$ is generally higher than $t=8830t_g$. Similar conclusions can be drawn after comparing the brightness of the images with the ones previously shown. The reasons for these observations become clearer when we examine the variation of dimensionless magnetic flux and accretion rate as shown in Fig.~\ref{fig:mag_flux}. Firstly, when the magnetic flux reaches its peak, more plasma is expelled by the magnetic pressure, resulting in a sudden decrease in the accretion rate.  A stronger outflow results in stronger emissions, explaining why the snapshots at $6020t_g$ are considerably brighter. Moreover, the intensity of thermal synchrotron emission is positively correlated with magnetic intensity. Within the saturation regime, the magnetic flux near the event horizon remains near its maximum (indicating that the black hole retains more magnetic field lines), which results in stronger emissions due to the higher magnetic intensity.

\section{Summary and Discussion}

\label{sec5}

Utilizing 3D GRMHD simulations, we have successfully generated  a geometrically thin jet in an accretion system surrounding a Kerr BH by imposing a strong magnetic field. We have examined various aspects of the jet's images and identified their features as well. 

In our study, we categorized the evolution of the entire accretion system into three primary regimes by identifying different variation behaviors of both the accretion rate and the dimensionless magnetic flux crossing the event horizon. In the last two regimes we primarily analyzed, referred to as the saturation regime and the MAD regime, the thin jet forms when the ratio of magnetic pressure to gas pressure reaches distinct maxima near the jet surface. As we have elucidated, within the saturation regime, the entire system remains in an unstable equilibrium when gravity is counterbalanced by the magnetic field. This equilibrium is disrupted by  over-density and under-density fluctuations in the accretion flow caused by MRI.

We have plotted both time-averaged images  and snapshots at two specific moments (related to the maximal and minimal dimensionless magnetic flux within the MAD regime we simulated). We found that, although the snapshots exhibit obvious asymmetry due to chaotic oscillation within the MAD regime, the time-averaged images of the two regimes are nearly identical. This finding suggests that we can discern some features of an unstable accretion system from stationary approximations when considering the images.

We have also displayed the observed total fluxes for different regimes and snapshots. From these, we concluded that our model is insensitive to frequency and that the total intensity generally correlates positively with the strength of the dimensionless magnetic flux, a typical feature of synchrotron emission.

The most intriguing feature we have identified is the U-shaped brighter lines, which are particularly noticeable in the images observed from $17^{\circ}$ and $163^{\circ}$. This is not a consequence of the beaming effect or  gravitational lensing. Instead, these brighter lines originate from the light rays that skim over the jet surface, namely, the primary emission region. The presence of the U-shaped brighter lines is a unique feature of geometrically and optically thin jets. We anticipate that a thinner surface would produce sharper and brighter lines, while these lines would disappear when the absorption becomes substantial or if the entire funnel region becomes the emission region.

There are numerous fascinating topics that deserve further investigations. Here we list some of them. Firstly, in order to simplify our numerical simulations, we primarily focused on a geometrically thin jet, with the accretion disk positioned at the equator. It would be interesting to relax these constraints and investigate the images of jets with various configurations.
Secondly, since both the disk and the jet contribute to the characteristics of the accretion system, particularly for compact radio cores such as supermassive BH in galaxies, it is of paramount importance to study the images of both the disk and the jet, and compare them with astronomical observations. 
Moreover, in the case of M87*, a triple-ridge jet structure was observed on larger scales ($\gtrsim 100 R_s$) \cite{Lu:2023bbn}, and the position of the central ridge implies the existence of a central spine originating from the ring's center. However, we did not observe such phenomena  in our framework. It would be highly interesting to investigate how this triple-ridge structure emerges in GRMHD simulations. Such a study could significantly advance our understanding of BH accretion and jet formation in the near future.

\section*{Acknowledgments}

Ye Shen thanks Rong Du for useful discussions about GRMHD. The work is partly supported by NSFC Grant No. 12275004, 12205013 and 11873044. Z.Y. Fan was supported in part by the National Natural Science Foundations of China with Grant No. 11805041 and No. 11873025. MG is also endorsed by "the Fundamental Research Funds for the Central Universities" with Grant No. 2021NTST13.

\appendix
\label{app}
\section{Rescaling}
\label{sec:rescale}

In GRMHD, we set $G=M=c=1$ so that all physical quantities are measured in these units, such as $r_g=GMc^{-2}$ to be a unit length and $Mr_g^{-3}=c^6G^{-3}M^{-2}$ to be a unit mass density. However, if we want to match our simulations with a real astronomical system, we need to turn back to the units usually adopted (SI or cgs).

Generally, for real accretion systems, the mass of plasma around the BH is varying (compared to the BH mass) so it is inefficient to simulate an evolution for a certain mass ratio. On the other hand, the mass of matter is much smaller than the BH mass in most of accretion system. When we set $M=1$, all physical values in magnetohydrodynamics would be infinitesimal which causes huge numerical imprecision. Fortunately, for a linear dependent theory, the evolution will be rescaling free \cite{PATOKA}, which is the case in GRMHD. Explicitly, if we rescale physical quantities by a constant $\lambda$ as
\begin{equation}
    \label{eq:rescale}
    \rho\rightarrow\lambda\rho,~~p\rightarrow\lambda p,~~b^{\mu}\rightarrow\lambda^{\frac{1}{2}}b^{\mu},~~u^{\mu}\rightarrow u^{\mu},
\end{equation}
then the evolution equations become
\begin{equation}
    \begin{array}{l}
        \lambda\nabla_{\mu}\left(\rho u^{\mu}\right)=0,
        \\
        \lambda\nabla_{\mu}T^{\mu\nu}=0,
        \\
        \lambda^{\frac{1}{2}}\nabla_{\mu}{}^\ast F^{\mu\nu}=0,
    \end{array}
    \label{eq:rescale}
\end{equation}
which are obviously equivalent to the equations in Eq.~\eqref{eq:conservation}. In this case, as long as the dynamics is linearly dependent (independent of viscosity, heat dissipation, radiation etc), we can simply work in a proper unit in numerical simulation and then rescale the results to the values we want according to (\ref{eq:rescale}). Notice that GRMHD cannot provide a temperature directly. The rescaling of temperature depends on the EoS we choose. If we regard the fluid as ideal gas, whose temperature is proportional to the ratio of pressure to mass density, the temperature is invariant after rescaling. However, the emissivity of thermal radiation depends on both the particle density and the magnetic field, the image and the spectrum of the jet are varied after rescaling.

For M87*, the accretion system is well described by radiatively inefficient accretion flow (RIAF) \cite{EventHorizonTelescope:2021srq}. It implies that the above linear equations of motion suitably describes the accretion of M87* to a certain extent. The recommended accretion rate of M87* is $\sim 10^{-3}M_{\odot}yr^{-1}$ (see \cite{AccretionRate}) so that we take the rescaling index $\lambda$, from numerical scale to M87*, to be $\lambda\sim10^{-15}$.

\section{Results of Curve Fitting}
\label{sec:result_curve_fit}

Here we show the results of curve fitting. We use a power function $f=Cr^A$ to fit the numerical results as an empirical choice. As a comparison, we also apply an exponential function $f=Ce^{Ar}$ to fit the results. Because the variation of $\Delta\theta_{\rm jet}$ along $r$ is chaotic, it is not suitable to describe it by a smooth function. While the range of its variation is not big ($\lesssim0.001$), we just pick the mean value (0.027) to be a uniform thickness of emission region when plotting images.

We need the equation of state to determine the gas temperature. Taking the fluid as classical ideal gas, we have
\be p=\frac{\rho}{\mu}k_BT \,,\ee
where $\mu$ is the average particle mass.

\begin{figure}
    \begin{subfigure}[b]{0.5\textwidth}
        \centering
        \includegraphics[width=\textwidth]{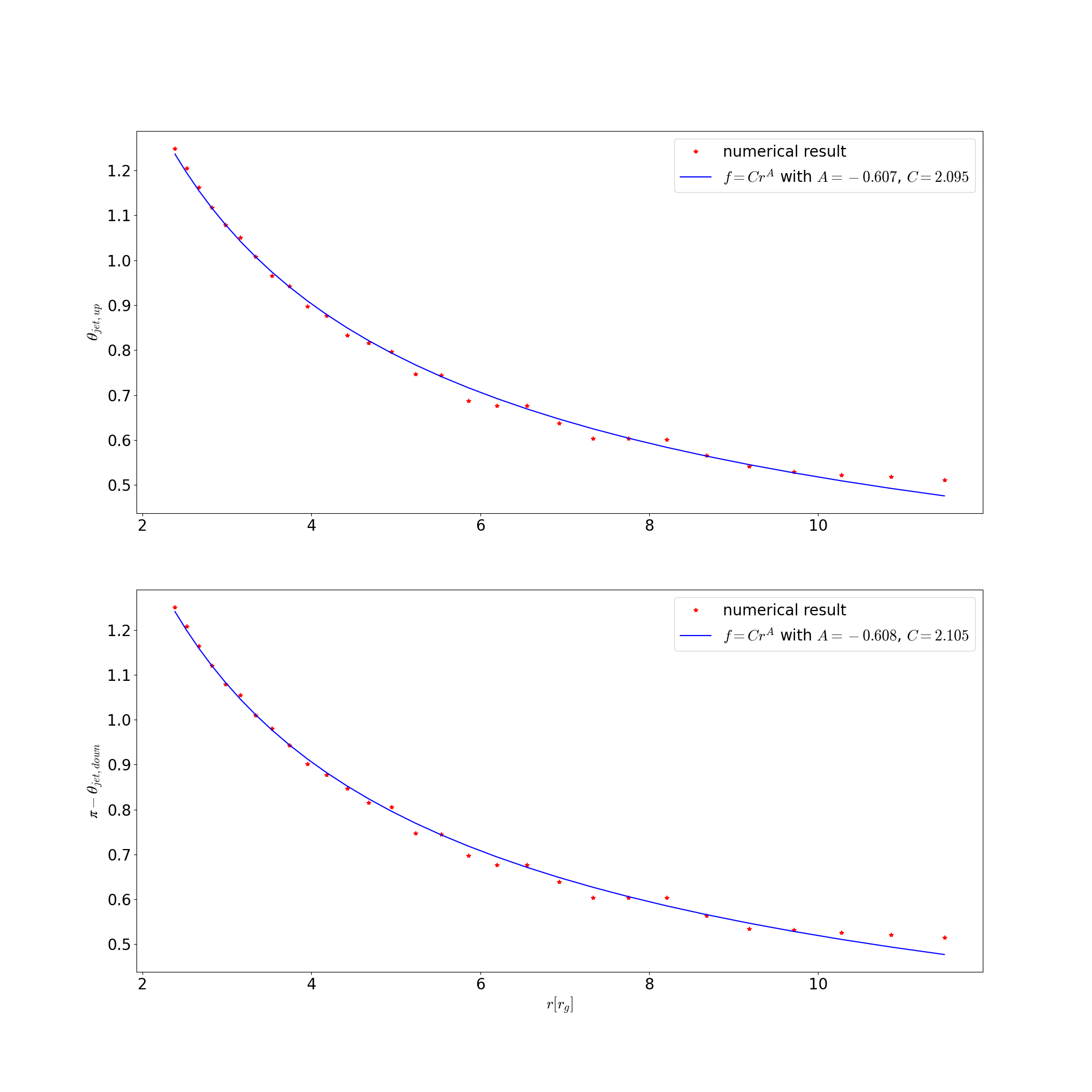}
        \caption{2000-4000$t_{g}$ averaged}
        \label{the_ave2-4}
    \end{subfigure}
    \begin{subfigure}[b]{0.5\textwidth}
        \centering
        \includegraphics[width=\textwidth]{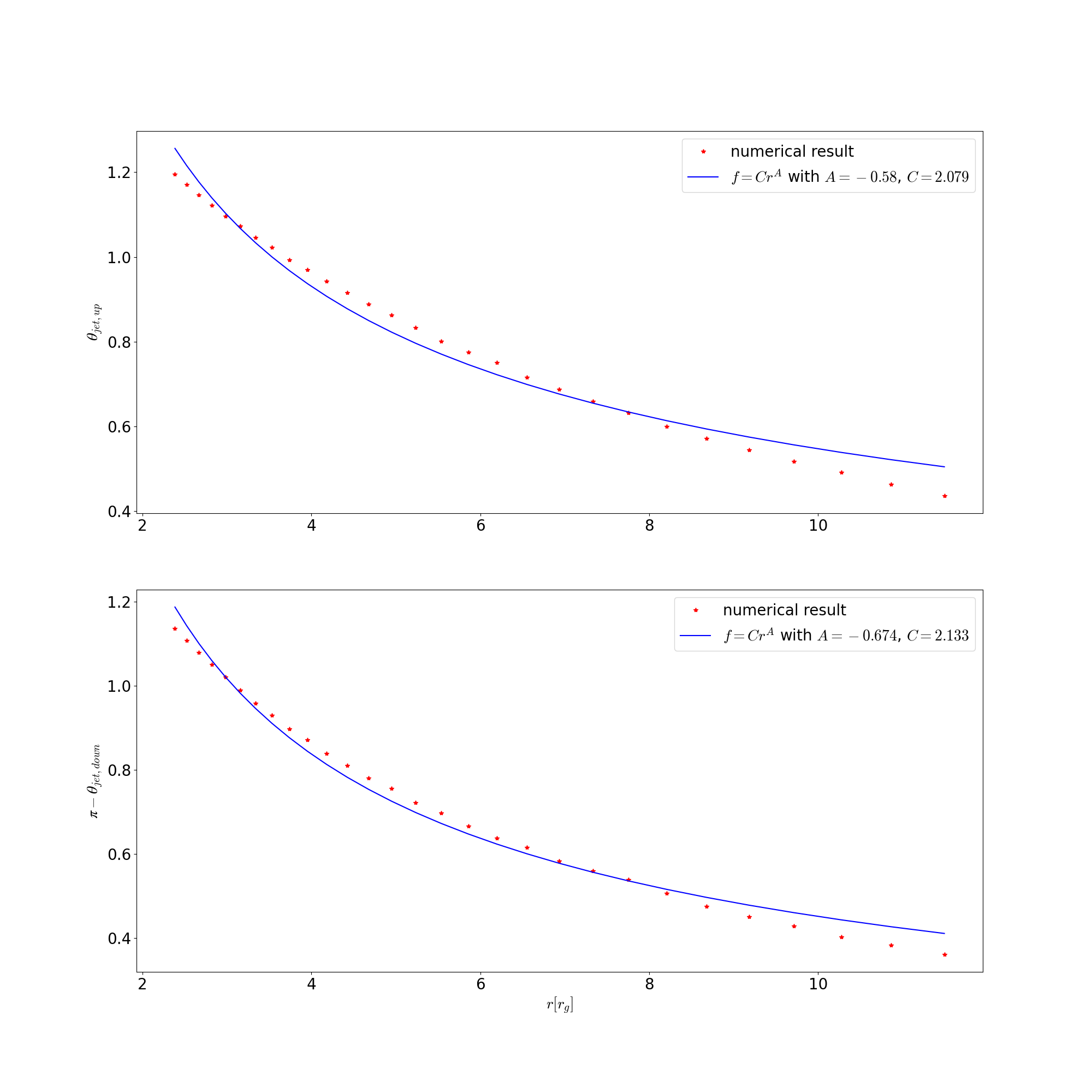}
        \caption{4000-10000$t_{g}$ averaged}
        \label{the_ave4-10}
    \end{subfigure}
    \begin{subfigure}[b]{0.5\textwidth}
        \centering
        \includegraphics[width=\textwidth]{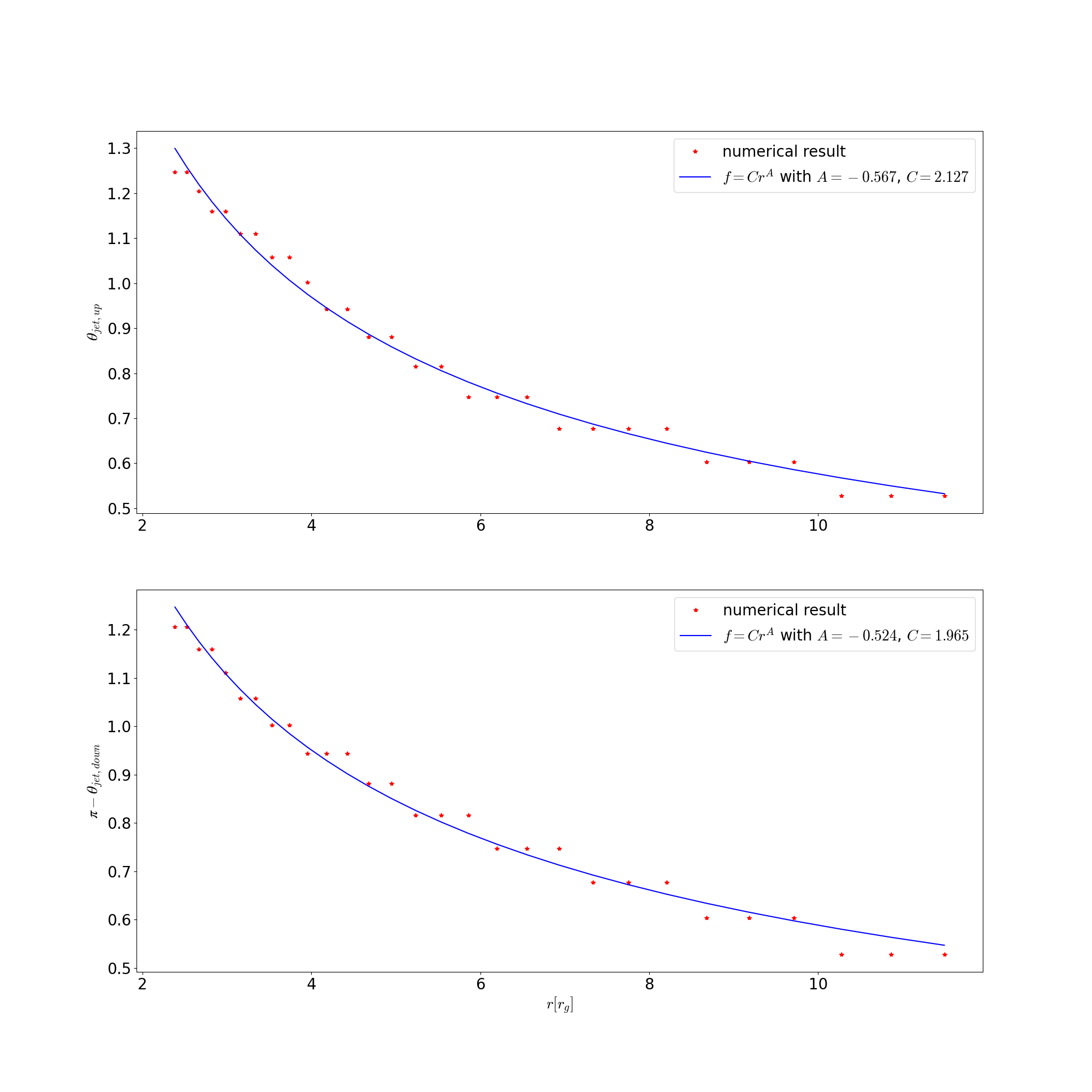}
        \caption{snapshot on 6020$t_{g}$}
        \label{the_snap602}
    \end{subfigure}
    \begin{subfigure}[b]{0.5\textwidth}
        \centering
        \includegraphics[width=\textwidth]{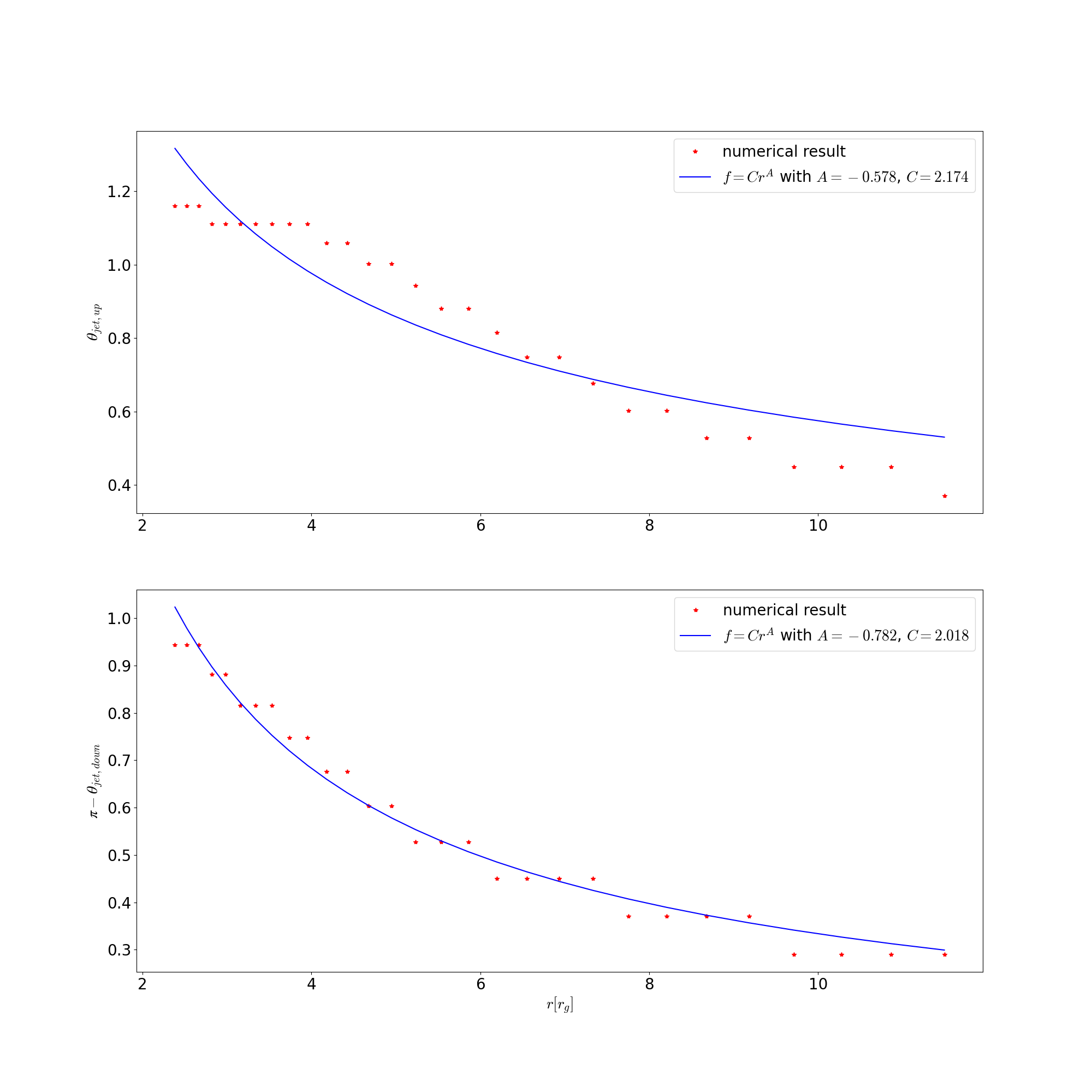}
        \caption{snapshot on 8830$t_{g}$}
        \label{the_snap883}
    \end{subfigure}	
    \caption{Curve fitting results of $\theta_{\rm jet}$.}
    \label{the}
\end{figure}

\begin{figure}
    \begin{subfigure}[b]{0.5\textwidth}
        \centering
        \includegraphics[width=\textwidth]{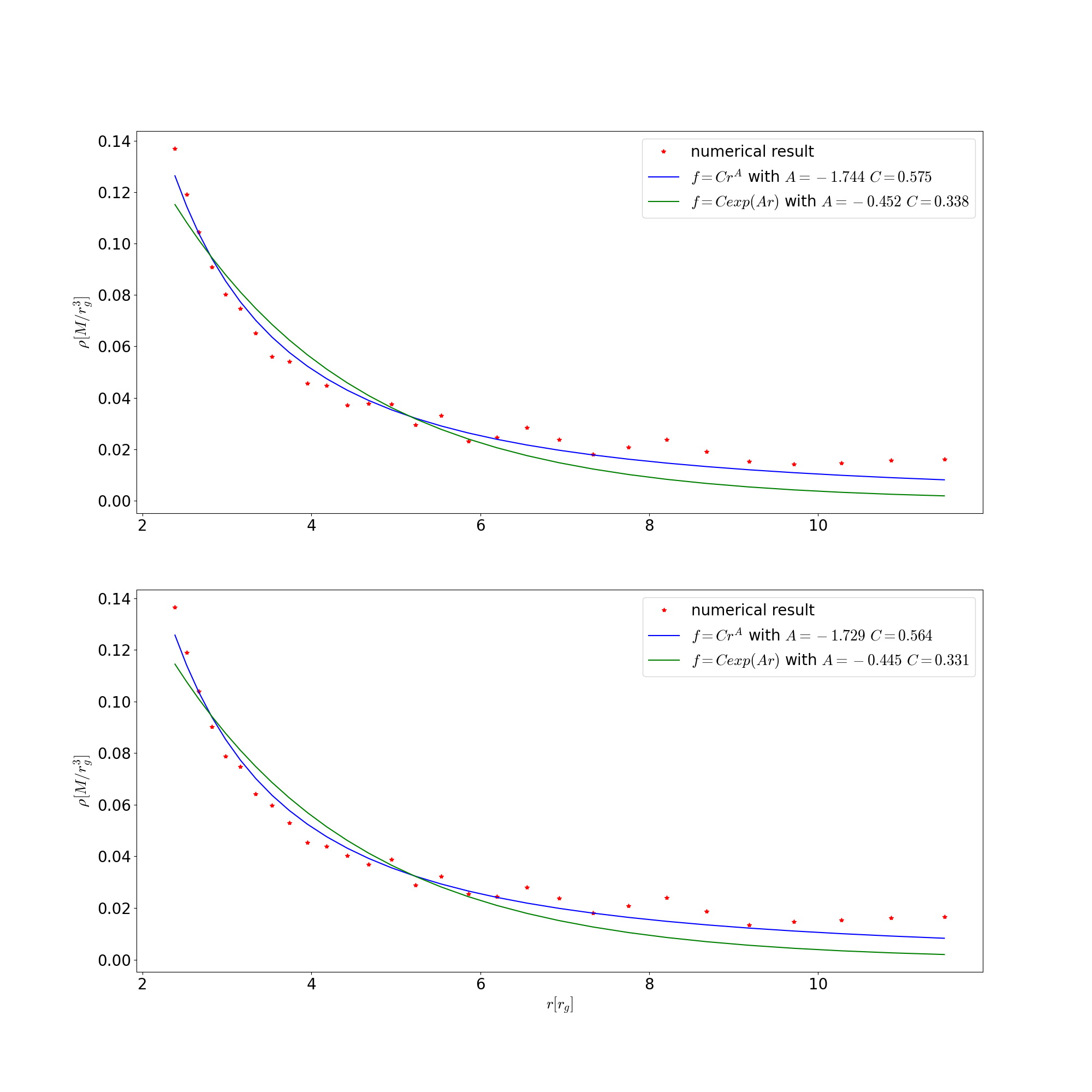}
        \caption{2000-4000$t_{g}$ averaged}
        \label{rho_ave2-4}
    \end{subfigure}
    \begin{subfigure}[b]{0.5\textwidth}
        \centering
        \includegraphics[width=\textwidth]{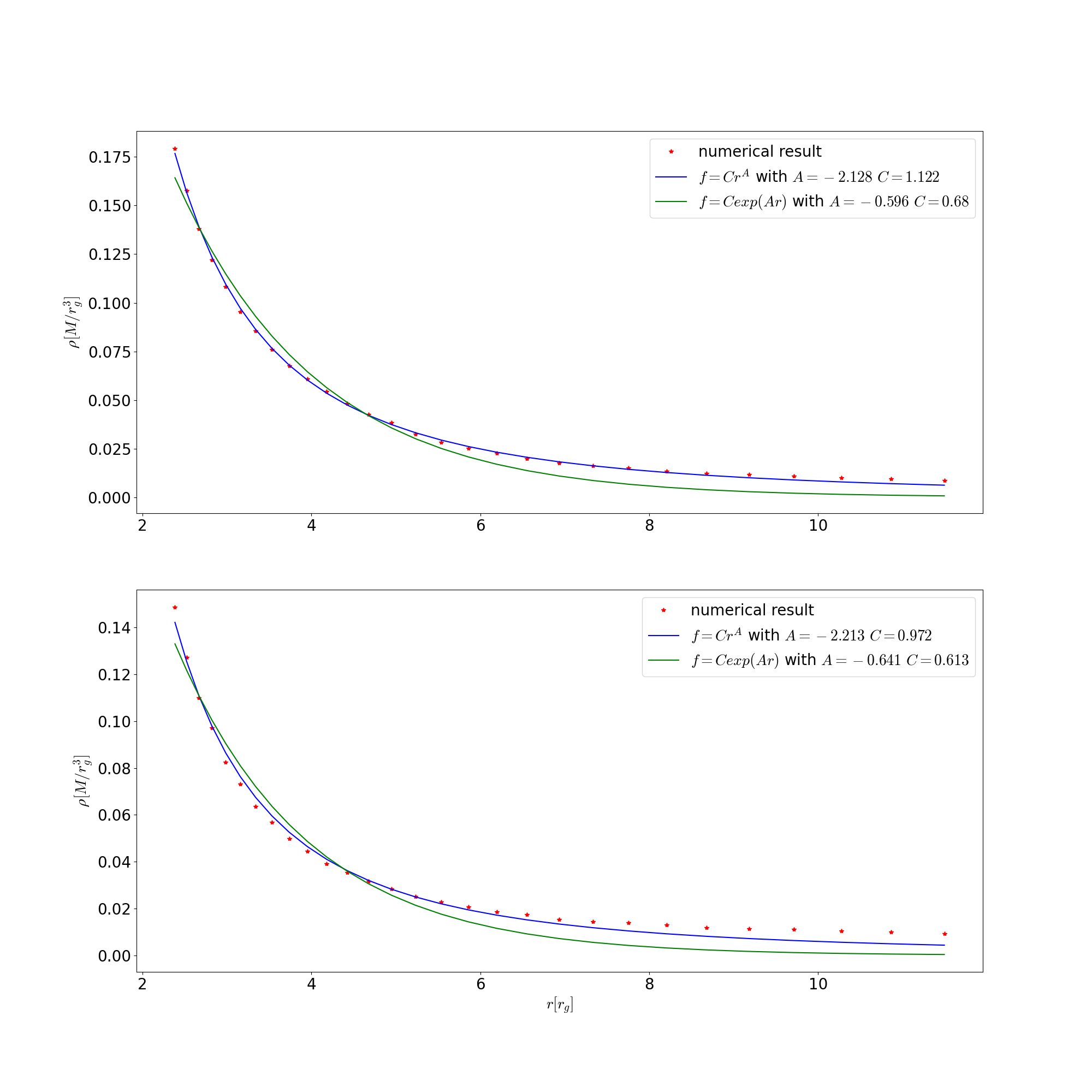}
        \caption{4000-10000$t_{g}$ averaged}
        \label{rho_ave4-10}
    \end{subfigure}
    \begin{subfigure}[b]{0.5\textwidth}
        \centering
        \includegraphics[width=\textwidth]{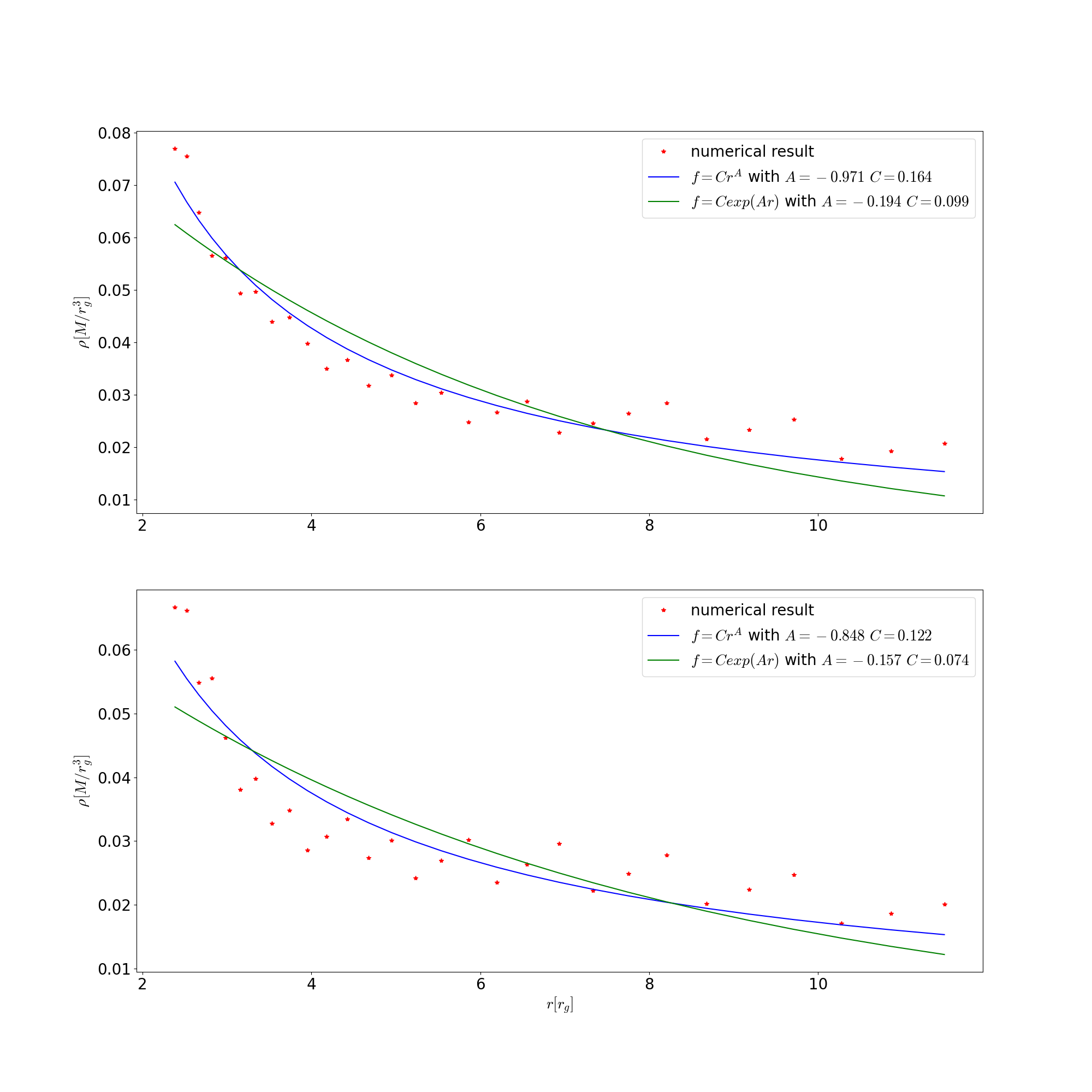}
        \caption{snapshot on 6020$t_{g}$}
        \label{rho_snap602}
    \end{subfigure}
    \begin{subfigure}[b]{0.5\textwidth}
        \centering
        \includegraphics[width=\textwidth]{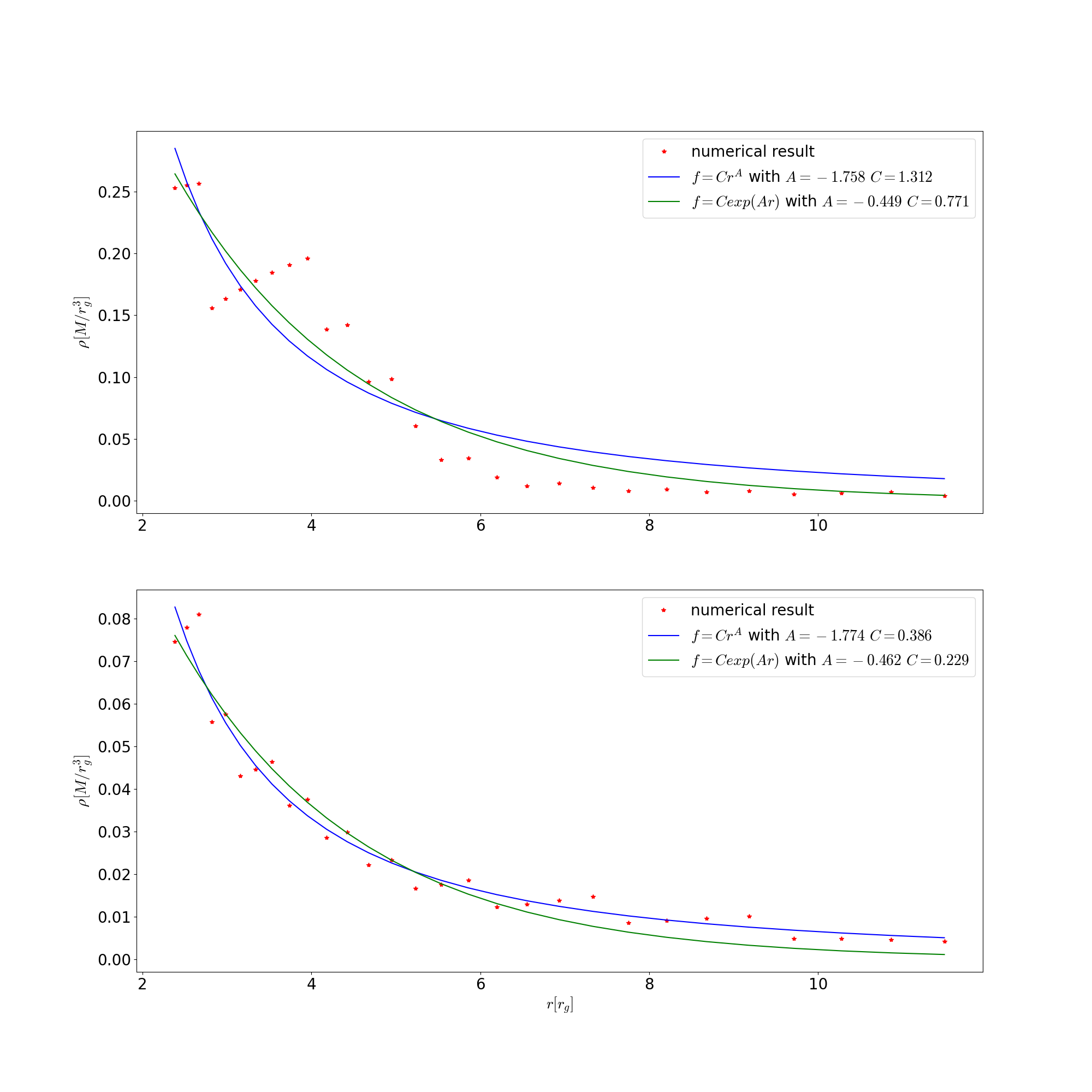}
        \caption{snapshot on 8830$t_{g}$}
        \label{rho_snap883}
    \end{subfigure}
    \caption{Curve fitting results of mass density.}
    \label{rho}
\end{figure}

\begin{figure}
    \begin{subfigure}[b]{0.5\textwidth}
        \centering
        \includegraphics[width=\textwidth]{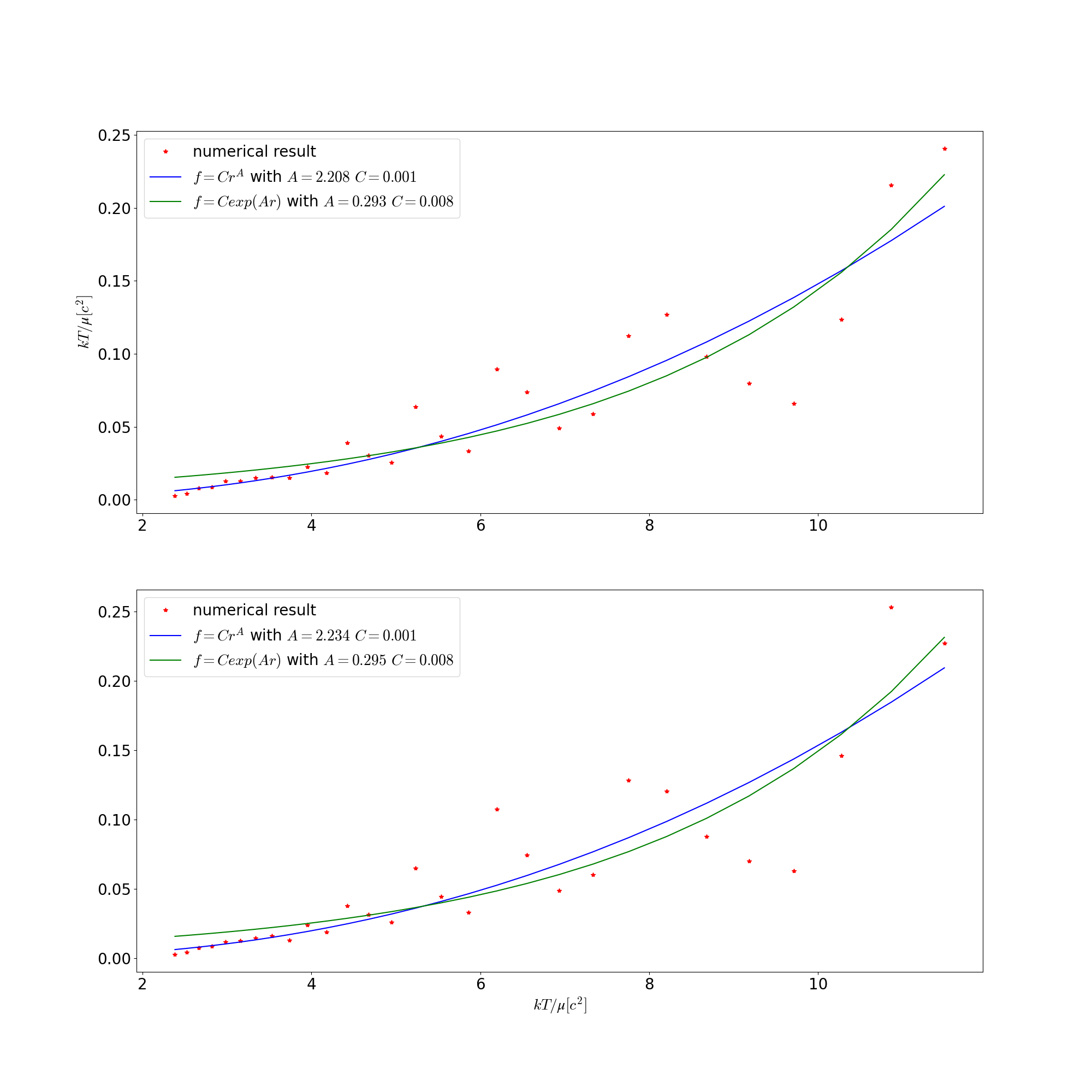}
        \caption{2000-4000$t_{g}$ averaged}
        \label{kT_ave2-4}
    \end{subfigure}
    \begin{subfigure}[b]{0.5\textwidth}
        \centering
        \includegraphics[width=\textwidth]{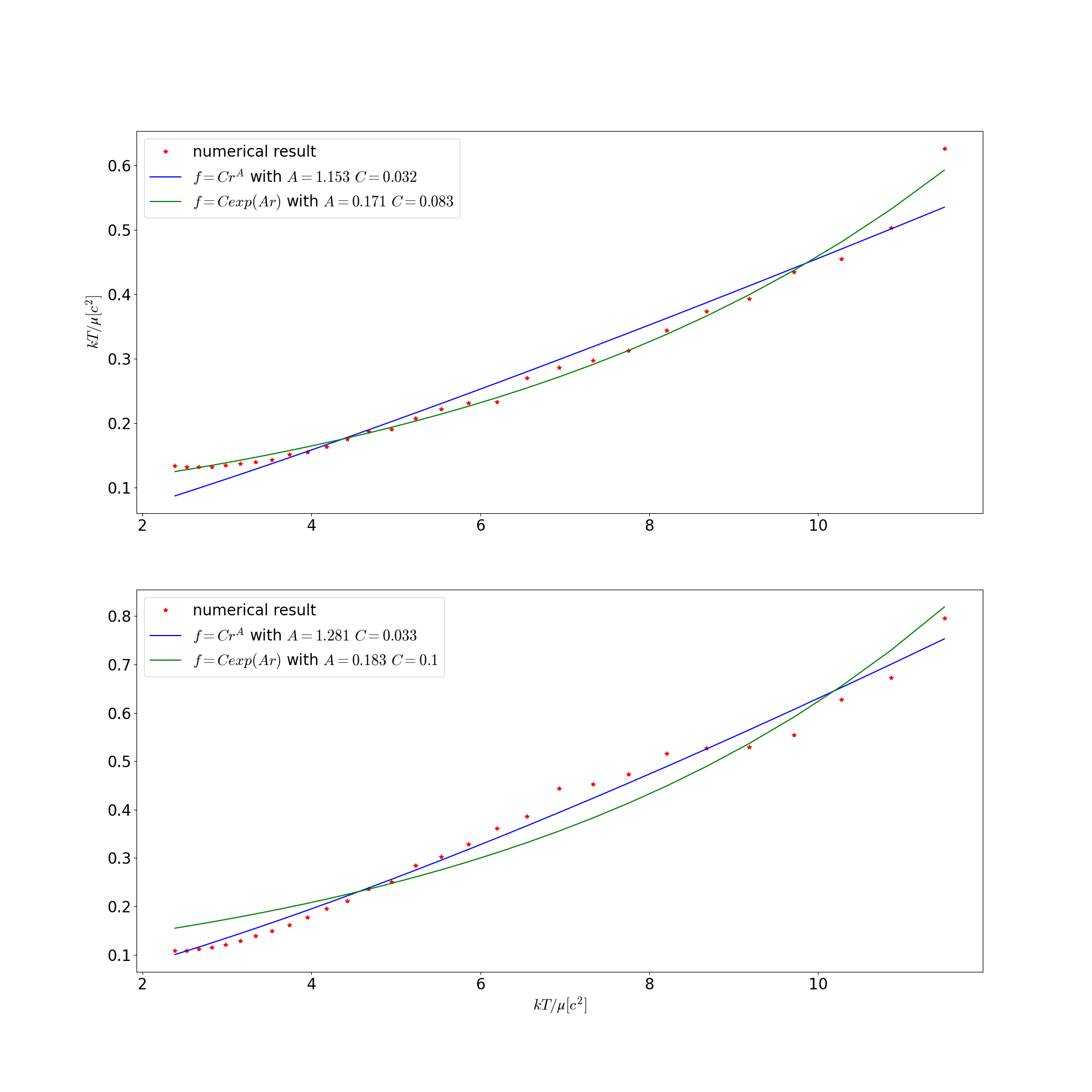}
        \caption{4000-10000$t_{g}$ averaged}
        \label{kT_ave4-10}
    \end{subfigure}
    \begin{subfigure}[b]{0.5\textwidth}
        \centering
        \includegraphics[width=\textwidth]{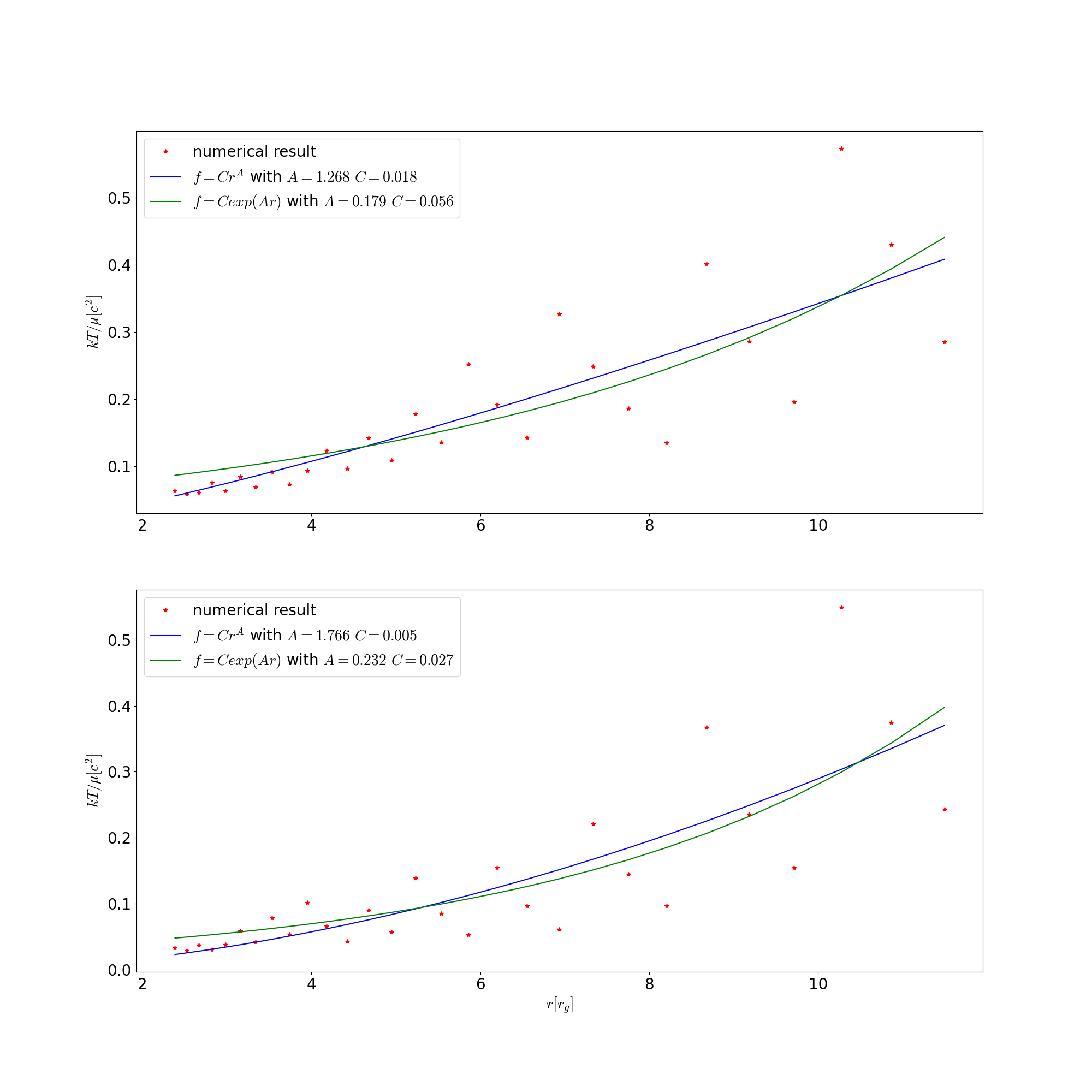}
        \caption{snapshot on 6020$t_{g}$}
        \label{kT_snap602}
    \end{subfigure}
    \begin{subfigure}[b]{0.5\textwidth}
        \centering
        \includegraphics[width=\textwidth]{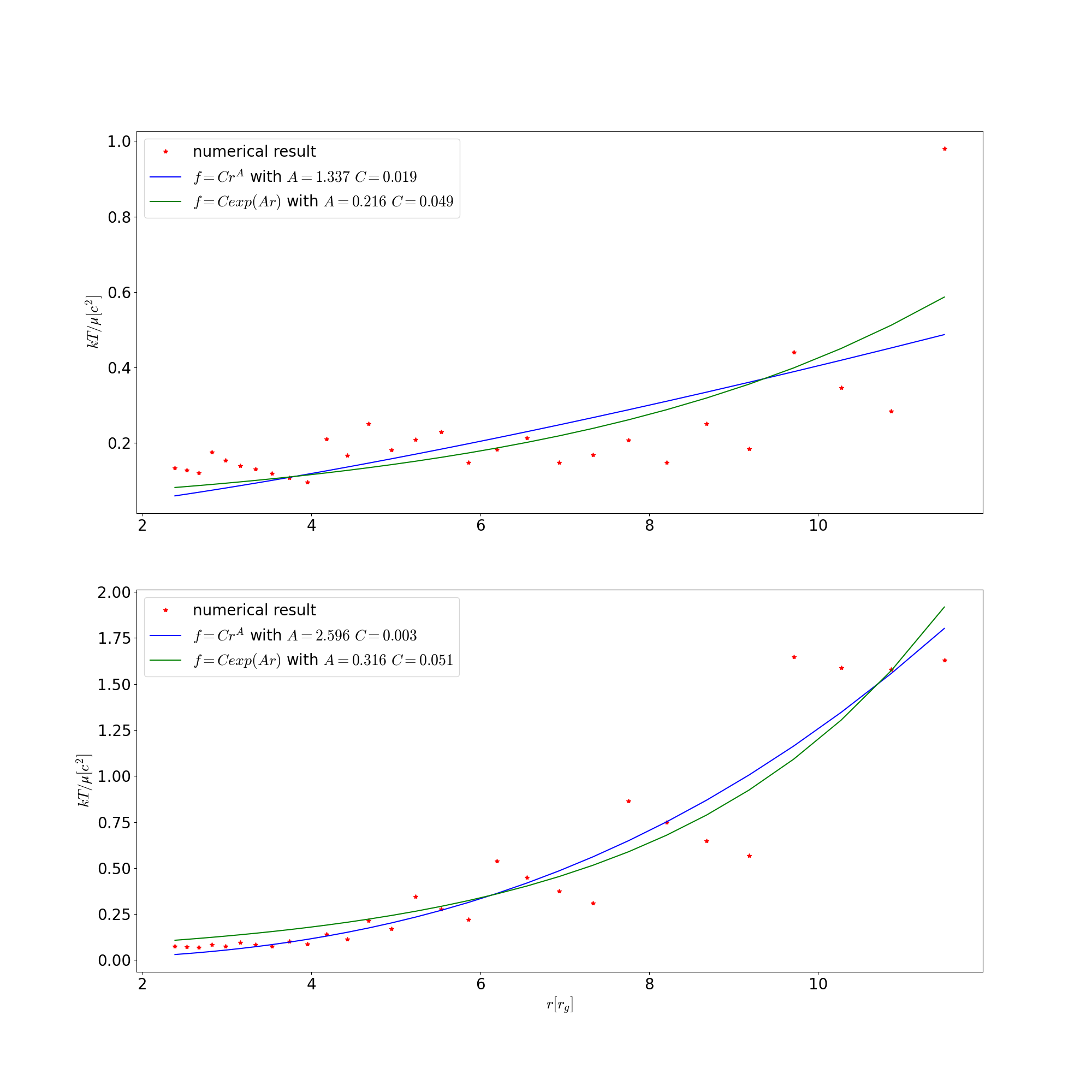}
        \caption{snapshot on 8830$t_{g}$}
        \label{kT_snap883}
    \end{subfigure}
    \caption{Curve fitting results of temperature.}
    \label{temp}
\end{figure}

We show the exact results of curve fitting  in Tab.~\ref{tab:jet_ave} and \ref{tab:jet_snap} for time-averages and snapshots respectively, where we provide both optimal values and standard deviations for all parameters. We also plot some of our curve fitting results in Fig.~\ref{the}, \ref{rho} and \ref{temp}.

\begin{table}[!ht]
    \centering
    \caption{Time Averaged Curve Fitting Results of Jet}
        \begin{tabular}{|c|c|c|c|c|c|}
            \hline
            \multicolumn{6}{|c|}{2000-4000$t_g$ time averaged} \\ \hline
            \multicolumn{2}{|c|}{} & \multicolumn{2}{c|}{North} & \multicolumn{2}{c|}{South} \\ \cline{3-6}
            \multicolumn{2}{|c|}{} & A & C & A & C \\ \hline
            \multirow{2}{*}{$\theta_{\rm jet}$($\pi-\theta_{\rm jet}$)} & optimal value & -0.607 & 2.09 & -0.608 & 2.10 \\ \cline{2-6}
            \multirow{2}{*}{} & standard deviation & $7.51\times10^{-3}$ & $2.31\times10^{-2}$ & $7.43\times10^{-3}$ & $2.29\times10^{-2}$ \\ \hline
            \multirow{2}{*}{$\rho$} & optimal value & -1.74 & 0.575 & -1.73 & 0.564 \\ \cline{2-6}
            \multirow{2}{*}{} & standard deviation & $6.57\times10^{-2}$ & $4.36\times10^{-2}$ & $6.43\times10^{-2}$ & $4.20\times10^{-2}$ \\ \hline
            \multirow{2}{*}{$\frac{kT}{\mu}$} & optimal value & 2.21 & $9.16\times10^{-4}$ & 2.23 & $8.94\times10^{-4}$ \\ \cline{2-6}
            \multirow{2}{*}{} & standard deviation & 0.261 & $5.40\times10^{-4}$ & $0.300$ & $6.05\times10^{-4}$ \\ \hline
            \multirow{2}{*}{$u^{r}$} & optimal value & -1.75 & -0.823 & -1.74 & -0.816 \\ \cline{2-6}
            \multirow{2}{*}{} & standard deviation & $3.70\times10^{-2}$ & $3.51\times10^{-2}$ & $3.59\times10^{-2}$ & $3.39\times10^{-2}$ \\ \hline
            \multirow{2}{*}{$u^{\theta}$} & optimal value & -1.98 & 0.197 & -1.97 & -0.194 \\ \cline{2-6}
            \multirow{2}{*}{} & standard deviation & $4.18\times10^{-2}$ & $9.19\times10^{-3}$ & $4.18\times10^{-2}$ & $9.09\times10^{-3}$ \\ \hline
            \multirow{2}{*}{$u^{\phi}$} & optimal value & -1.24 & 1.26 & -1.24 & 1.26 \\ \cline{2-6}
            \multirow{2}{*}{} & standard deviation & $2.18\times10^{-2}$ & $3.46\times10^{-2}$ & $2.22\times10^{-2}$ & $3.52\times10^{-2}$ \\ \hline
            \multirow{2}{*}{$B^{r}$} & optimal value & -2.01 & 0.936 & -2.01 & -0.936 \\ \cline{2-6}
            \multirow{2}{*}{} & standard deviation & $1.14\times10^{-2}$ & $1.19\times10^{-2}$ & $1.06\times10^{-2}$ & $1.11\times10^{-2}$ \\ \hline
            \multirow{2}{*}{$B^{\theta}$} & optimal value & -2.28 & -0.232 & -2.28 & $-0.231$ \\ \cline{2-6}
            \multirow{2}{*}{} & standard deviation & $1.60\times10^{-2}$ & $4.01\times10^{-3}$ & $1.57\times10^{-2}$ & $3.92\times10^{-3}$ \\ \hline
            \multirow{2}{*}{$B^{\phi}$} & optimal value & -1.79 & -1.77 & -1.79 & 1.77 \\ \cline{2-6}
            \multirow{2}{*}{} & standard deviation & $4.57\times10^{-2}$ & $9.27\times10^{-2}$ & $4.63\times10^{-2}$ & $9.41\times10^{-2}$ \\ \hline
            \hline
            \multicolumn{6}{|c|}{4000-10000$t_g$ time averaged} \\ \hline
            \multicolumn{2}{|c|}{} & \multicolumn{2}{c|}{North} & \multicolumn{2}{c|}{South} \\ \cline{3-6}
            \multicolumn{2}{|c|}{} & A & C & A & C \\ \hline
            \multirow{2}{*}{$\theta_{\rm jet}$($\pi-\theta_{\rm jet}$)} & optimal value & -0.580 & 2.08 & -0.674 & 2.13 \\ \cline{2-6}
            \multirow{2}{*}{} & standard deviation & $1.76\times10^{-2}$ & $5.40\times10^{-2}$ & $1.55\times10^{-2}$ & $4.77\times10^{-2}$ \\ \hline
            \multirow{2}{*}{$\rho$} & optimal value & -2.13 & 1.12 & -2.21 & 0.972 \\ \cline{2-6}
            \multirow{2}{*}{} & standard deviation & $1.53\times10^{-2}$ & $1.88\times10^{-2}$ & $4.87\times10^{-2}$ & $5.15\times10^{-2}$ \\ \hline
            \multirow{2}{*}{$\frac{kT}{\mu}$} & optimal value & 1.15 & $3.21\times10^{-2}$ & 2.18 & $3.30\times10^{-2}$ \\ \cline{2-6}
            \multirow{2}{*}{} & standard deviation & $4.97\times10^{-2}$ & $3.35\times10^{-3}$ & $3.43\times10^{-2}$ & $2.41\times10^{-3}$ \\ \hline
            \multirow{2}{*}{$u^{r}$} & optimal value & -1.84 & -0.744 & -1.89 & -0.871 \\ \cline{2-6}
            \multirow{2}{*}{} & standard deviation & $1.79\times10^{-2}$ & $1.52\times10^{-2}$ & $2.32\times10^{-2}$ & $2.29\times10^{-2}$ \\ \hline
            \multirow{2}{*}{$u^{\theta}$} & optimal value & -1.71 & $9.35\times10^{-2}$ & -1.75 & $-9.87\times10^{-2}$ \\ \cline{2-6}
            \multirow{2}{*}{} & standard deviation & $4.76\times10^{-2}$ & $5.16\times10^{-3}$ & $5.47\times10^{-2}$ & $6.23\times10^{-3}$ \\ \hline
            \multirow{2}{*}{$u^{\phi}$} & optimal value & -1.30 & 1.07 & -1.43 & 1.27 \\ \cline{2-6}
            \multirow{2}{*}{} & standard deviation & $1.67\times10^{-2}$ & $2.23\times10^{-2}$ & $9.47\times10^{-3}$ & $1.46\times10^{-2}$ \\ \hline
            \multirow{2}{*}{$B^{r}$} & optimal value & -2.21 & 1.01 & -2.02 & -0.938 \\ \cline{2-6}
            \multirow{2}{*}{} & standard deviation & $2.71\times10^{-2}$ & $2.98\times10^{-2}$ & $2.15\times10^{-2}$ & $2.25\times10^{-2}$ \\ \hline
            \multirow{2}{*}{$B^{\theta}$} & optimal value & -2.07 & -0.124 & -2.05 & -0.124 \\ \cline{2-6}
            \multirow{2}{*}{} & standard deviation & $2.40\times10^{-2}$ & $3.31\times10^{-3}$ & $3.70\times10^{-2}$ & $5.07\times10^{-3}$ \\ \hline
            \multirow{2}{*}{$B^{\phi}$} & optimal value & -1.77 & -1.53 & -1.48 & 1.19 \\ \cline{2-6}
            \multirow{2}{*}{} & standard deviation & $5.09\times10^{-2}$ & $8.95\times10^{-2}$ & $6.13\times10^{-2}$ & $8.81\times10^{-2}$ \\ \hline
        \end{tabular}
    \label{tab:jet_ave}   
\end{table}

\begin{table}[!ht]
    \centering
    \caption{Snapshot Curve Fitting Results of Jet}
        \begin{tabular}{|c|c|c|c|c|c|}
            \hline
            \multicolumn{6}{|c|}{6020$t_g$ snapshot} \\ \hline
            \multicolumn{2}{|c|}{} & \multicolumn{2}{c|}{North} & \multicolumn{2}{c|}{South} \\ \cline{3-6}
            \multicolumn{2}{|c|}{} & A & C & A & C \\ \hline
            \multirow{2}{*}{$\theta_{\rm jet}$($\pi-\theta_{\rm jet}$)} & optimal value & -0.567 & 2.13 & -0.524 & 1.96 \\ \cline{2-6}
            \multirow{2}{*}{} & standard deviation & $1.21\times10^{-2}$ & $3.80\times10^{-2}$ & $1.17\times10^{-2}$ & $3.45\times10^{-2}$ \\ \hline
            \multirow{2}{*}{$\rho$} & optimal value & -0.971 & 0.164 & -0.848 & 0.122 \\ \cline{2-6}
            \multirow{2}{*}{} & standard deviation & $5.19\times10^{-2}$ & $1.14\times10^{-2}$ & $7.35\times10^{-2}$ & $1.23\times10^{-2}$ \\ \hline
            \multirow{2}{*}{$\frac{kT}{\mu}$} & optimal value & 1.27 & $1.85\times10^{-2}$ & 1.77 & $4.95\times10^{-3}$ \\ \cline{2-6}
            \multirow{2}{*}{} & standard deviation & 0.198 & $7.79\times10^{-3}$ & 0.321 & $3.51\times10^{-3}$ \\ \hline
            \multirow{2}{*}{$u^{r}$} & optimal value & -1.65 & -0.729 & -1.63 & -0.774 \\ \cline{2-6}
            \multirow{2}{*}{} & standard deviation & $2.55\times10^{-2}$ & $2.18\times10^{-2}$ & $2.51\times10^{-2}$ & $2.28\times10^{-2}$ \\ \hline
            \multirow{2}{*}{$u^{\theta}$} & optimal value & -1.52 & 0.120 & -1.46 & -0.119 \\ \cline{2-6}
            \multirow{2}{*}{} & standard deviation & $9.34\times10^{-2}$ & $1.34\times10^{-2}$ & $8.71\times10^{-2}$ & $1.26\times10^{-3}$ \\ \hline
            \multirow{2}{*}{$u^{\phi}$} & optimal value & -1.25 & 1.25 & -1.28 & 1.33 \\ \cline{2-6}
            \multirow{2}{*}{} & standard deviation & $1.60\times10^{-2}$ & $2.51\times10^{-2}$ & $1.82\times10^{-2}$ & $3.04\times10^{-2}$ \\ \hline
            \multirow{2}{*}{$B^{r}$} & optimal value & -2.20 & 0.884 & -2.18 & -0.918 \\ \cline{2-6}
            \multirow{2}{*}{} & standard deviation & $3.94\times10^{-2}$ & $3.80\times10^{-2}$ & $2.87\times10^{-2}$ & $2.88\times10^{-2}$ \\ \hline
            \multirow{2}{*}{$B^{\theta}$} & optimal value & -1.92 & -0.122 & -1.85 & -0.118 \\ \cline{2-6}
            \multirow{2}{*}{} & standard deviation & $5.73\times10^{-2}$ & $7.85\times10^{-3}$ & $6.68\times10^{-2}$ & $9.11\times10^{-3}$ \\ \hline
            \multirow{2}{*}{$B^{\phi}$} & optimal value & -1.88 & -1.77 & -1.89 & 1.81 \\ \cline{2-6}
            \multirow{2}{*}{} & standard deviation & $5.20\times10^{-2}$ & 0.104 & $4.93\times10^{-2}$ & 0.101 \\ \hline
            \hline
            \multicolumn{6}{|c|}{8830$t_g$ snapshot} \\ \hline
            \multicolumn{2}{|c|}{} & \multicolumn{2}{c|}{North} & \multicolumn{2}{c|}{South} \\ \cline{3-6}
            \multicolumn{2}{|c|}{} & A & C & A & C \\ \hline
            \multirow{2}{*}{$\theta_{\rm jet}$($\pi-\theta_{\rm jet}$)} & optimal value & -0.578 & 2.17 & -0.782 & 2.02 \\ \cline{2-6}
            \multirow{2}{*}{} & standard deviation & $4.72\times10^{-2}$ & 0.152 & $2.44\times10^{-2}$ & $6.90\times10^{-2}$ \\ \hline
            \multirow{2}{*}{$\rho$} & optimal value & -1.76 & 1.31 & -1.77 & 0.386 \\ \cline{2-6}
            \multirow{2}{*}{} & standard deviation & 0.186 & 0.281 & $7.98\times10^{-2}$ & $3.54\times10^{-2}$ \\ \hline
            \multirow{2}{*}{$\frac{kT}{\mu}$} & optimal value & 1.34 & $1.86\times10^{-2}$ & 2.60 & $3.18\times10^{-3}$ \\ \cline{2-6}
            \multirow{2}{*}{} & standard deviation & 0.301 & $1.20\times10^{-2}$ & 0.245 & $1.78\times10^{-3}$ \\ \hline
            \multirow{2}{*}{$u^{r}$} & optimal value & -2.21 & -0.820 & -2.05 & -1.14 \\ \cline{2-6}
            \multirow{2}{*}{} & standard deviation & $7.14\times10^{-2}$ & $6.38\times10^{-2}$ & $5.14\times10^{-2}$ & $6.49\times10^{-2}$ \\ \hline
            \multirow{2}{*}{$u^{\theta}$} & optimal value & -1.76 & $6.25\times10^{-2}$ & -2.03 & -0.105 \\ \cline{2-6}
            \multirow{2}{*}{} & standard deviation & $8.63\times10^{-2}$ & $6.21\times10^{-3}$ & $8.58\times10^{-2}$ & $9.99\times10^{-3}$ \\ \hline
            \multirow{2}{*}{$u^{\phi}$} & optimal value & -1.41 & 1.10 & -1.61 & 1.58 \\ \cline{2-6}
            \multirow{2}{*}{} & standard deviation & $4.91\times10^{-2}$ & $6.58\times10^{-2}$ & $7.97\times10^{-3}$ & 0.148 \\ \hline
            \multirow{2}{*}{$B^{r}$} & optimal value & -3.04 & 2.17 & -2.10 & -1.25 \\ \cline{2-6}
            \multirow{2}{*}{} & standard deviation & $3.81\times10^{-2}$ & $5.26\times10^{-2}$ & $7.38\times10^{-2}$ & $5.39\times10^{-3}$ \\ \hline
            \multirow{2}{*}{$B^{\theta}$} & optimal value & -1.86 & $-6.43\times10^{-2}$ & -2.18 & -0.125 \\ \cline{2-6}
            \multirow{2}{*}{} & standard deviation & $7.38\times10^{-2}$ & $5.39\times10^{-3}$ & $6.86\times10^{-2}$ & $9.40\times10^{-3}$ \\ \hline
            \multirow{2}{*}{$B^{\phi}$} & optimal value & -2.10 & -2.12 & -1.55 & 1.38 \\ \cline{2-6}
            \multirow{2}{*}{} & standard deviation & $8.80\times10^{-2}$ & 0.206 & $4.43\times10^{-2}$ & $7.30\times10^{-2}$ \\ \hline
        \end{tabular}
    \label{tab:jet_snap}   
\end{table}


\newpage
\bibliographystyle{utphys}
\bibliography{references}

\end{document}